\documentclass[a4paper]{amsart}
\usepackage{graphics}

\makeatletter

\providecommand{\LyX}{L\kern-.1667em\lower.25em\hbox{Y}\kern-.125emX\@}

\theoremstyle{plain}    
\newtheorem{thm}{Theorem} 
\theoremstyle{plain}    
\newtheorem{prop}{Proposition} 
\theoremstyle{definition}
\newtheorem{defn}{Definition}

\usepackage[T1]{fontenc}
\usepackage{a4}


\begin{document}

\title{Topological properties of quantum periodic Hamiltonians}

\author{Frédéric Faure }

\address{LPMMC (Maison des Magisteres Jean Perrin, CNRS) BP 166 38042 Grenoble Cedex
9 France \thanks{
email: faure@labs.polycnrs-gre.fr 
} . }

\begin{abstract}
We consider periodic quantum Hamiltonians on the torus phase space (Harper-like
Hamiltonians). We calculate the topological Chern index which characterizes
each spectral band in the generic case. This calculation is made by a semi-classical
approach with use of quasi-modes. As a result, the Chern index is equal to the
homotopy of the path of these quasi-modes on phase space as the Floquet parameter
\( \theta  \) of the band is varied. It is quite interesting that the Chern
indices, defined as topological quantum numbers, can be expressed from simple
properties of the classical trajectories.

P.A.C.S. number: 03.65.Sq, 73.50.Jt, 73.40.Hm, 73.20.Dx. key words: semi-classical
analysis, tunneling effect, topological numbers, Harper models.
\end{abstract}
\maketitle


\section{Introduction}

Topological quantum numbers are different from quantum numbers based on symmetry
because they are insensitive to the imperfections of the systems in which they
are observed. In some sense, topological properties are robust properties. Topological
quantum numbers have become very important in recent years in condensed matter
physics, where measurements of voltage and electrical resistance can be conveniently
expressed in terms of them. 

If you consider for example a simple model of non-interacting electrons moving
in a two-dimensional bi-periodic potential \( V(x,y) \) subject to a strong
uniform perpendicular magnetic field \( B_{z} \) and a low electrical field
\( E_{x} \), the Hall conductivity \( \sigma _{xy} \) of a given filled Landau
electronic band turns out to be proportional to an integer \( C \) \cite{iqhe2}:
\begin{equation}
\label{e:iqhe}
\sigma _{xy}=\frac{e^{2}}{h}C
\end{equation}
 \( C \) is the Chern index of the band, describing the topology of its fiber
bundle structure \cite{chern6}\cite{bellissard1}\cite{thouless1}\cite{chern4}\cite{chern3}.

We investigate in this paper the value of \( C \) as a function of the potential
\( V \). In the limit of high magnetic field \( B_{z} \), the above model
is mapped onto the well-known Harper model: the potential \( V \) is considered
as a perturbation of the cyclotron motion, and the averaging method of mechanics
gives an effective Hamiltonian equal to the average of \( V \) on the cyclotron
circles. We neglect the coupling between the Landau bands \cite{geisel1}. For
a high magnetic field, (hence for a small cyclotron radius,) this transformation
gives an effective Hamiltonian \( H_{eff}(q,p)\sim V(q,p) \), which is bi-periodic
in position and momentum (the phase space is a 2D torus), and an effective Planck
constant \( h_{eff}=hc/(eB) \). In this approximation, trajectories are the
contours of \( H_{eff}\sim V \). Furthermore, the expression of \( h_{eff} \)
shows that the high magnetic field regime corresponds to the semi-classical
limit. This model will be the starting point of our study in the next section.
For the sake of simplicity, we will denote \( \hbar _{eff} \) by \( \hbar  \)
in the sequel.

We will restrict ourself to the case where \( 1/h=N \) is an integer (but this
is not a strong restriction as discussed in section \ref{s:plane}). The spectrum
of these Harper-like models has then a finite band structure. To each band \( n=1\rightarrow N \)
is associated a topological Chern index \( C_{n} \). In this paper, we calculate
all these Chern indices in the semi-classical limit \( N\rightarrow \infty  \),
for a generic Hamiltonian \( H(q,p) \) on the torus.

The way we compute these topological indices is the following. First, we use
the fact that the classical dynamics generated by \( H \) is integrable, to
construct quasi-modes (the usual WKB construction) in section 4. An eigen-function
\( |\varphi _{n}(\theta _{1},\theta _{2})> \) of \( \hat{H} \) in band \( n \),
depends on two quantal parameters related to the periodicity conditions of the
wave function on the torus. A quasi-mode \( |\tilde{\psi }_{n}(\theta _{1},\theta _{2})> \)
is a quantum state localized on a trajectory which is very close to \( |\varphi _{n}(\theta _{1},\theta _{2})> \)
in the semi-classical limit \( N\rightarrow \infty  \),\cite{arnold1}\cite{lazutkin1}
(except when tunneling effect occurs).

Because of tunneling effect, the eigen-function \( |\varphi _{n}(\theta _{1},\theta _{2})> \)
jumps from one quasi-mode to another one as \( (\theta _{1},\theta _{2}) \)
are varying and follows a closed path on the torus. By treating correctly this
tunneling effects when it occurs, we obtain a good approximation of the bundle
of eigen-functions \( |\varphi _{n}(\theta _{1},\theta _{2})> \) by quasi modes
\( |\tilde{\psi }_{n}(\theta _{1},\theta _{2})> \), for every value of \( (\theta _{1},\theta _{2}) \).
This gives us the principal result of this paper: the Chern index \( C_{n} \)
is equal to the homotopy number \( I_{n} \) of this path, on the torus, see
eq.(\ref{e:result}). We obtain the Chern indices for the whole range of the
spectrum, and show that the non-contractible trajectories play a major role.

This result is based on properties presented in the appendix \ref{s:a2} specific
for the computation of the Chern indices. The analytical methods we use rest
on properties of zeros of the Bargmann representation (divisors of a holomorphic
section). This appendix is self-contained, and general results presented here
could be useful for other purposes where topological Chern indices are involved.

There is a constraint on the global spectrum \cite{chern3}\cite{iqhe1}\cite{chern4},
that \( \sum _{n}C_{n}=+1 \). In section 5, we give an explanation of this
in terms of classical trajectories. It is shown that the non-contractible trajectories
on the torus are responsible of this fact. This non-trivial total Chern number
is crucial for explanation of the Hall conductivity \cite{iqhe2}. A natural
question is then to determine which bands \( n \) have a non zero Chern index
\( C_{n} \) in the generic case? We answer this question in section \ref{s:generic},
where we explain how to compute the Chern indices of the whole spectrum, from
the classical Reeb graph.

Our approach is quite similar to Thouless et al.'s one \cite{iqhe2} where they
treated \( H_{\epsilon }(q,p)=cos(kq)+\epsilon cos(p) \). The difference is
that they treated \( H_{\epsilon } \) within a perturbation approximation (with
respect to \( \epsilon \ll 1 \)), whereas we use a semi-classical approach.
This enables us to treat any generic Hamiltonian \( H \), and to describe new
examples with non trivial \( C_{n} \), different from the Thouless et al.'s
one.

Our results extend previous work of the author \cite{fred1},\cite{fred2},
where the Chern indices were calculated for a restricted range of the spectrum
where the classical energy level \( \Sigma _{E}=\{(q,p)/H(q,p)=E\} \) consists
in two or three contractible trajectories. In these cases, it was shown that
\( C_{n} \) is zero except for special configurations of the trajectories.

We give a numerical illustration of these results in section 6. It is shown
that our results, although derived in the semi-classical limit \( N\rightarrow \infty  \),
are also valid for quite small \( N \). This illustrates the robustness of
the topological properties. This numerical example is also used throughout the
paper to confirm our calculations.

In section 8, we complete this work by giving a new (but equivalent) derivation
of the interpretation of the Chern index \( C \) for conductivity, more intuitive
than the usual one based on the Kubo formula \cite{iqhe2}: we show here that
\( C \) is equal to the mean velocity of a wave packet on the torus, when \( (\theta _{1},\theta _{2}) \)
move adiabatically.

\section{Classical trajectories of a generic Hamiltonian }

A generic classical Hamiltonian on the torus is a generic \( C^{\infty } \)
Morse function \( H(q,p) \) periodic in \( q\in I\! \! R \) and \( p\in I\! \! R \),
such that 
\begin{equation}
\label{e:periodic}
H(q+1,p)=H(q,p+1)=H(q,p).
\end{equation}
 In this section we investigate the topological properties of the trajectories
of \( H \), and introduce the Reeb graph \cite{toulet1}, which summarizes
them. The aim of this paper is to point out the rules that give the Chern indices
from the Reeb graph.

A trajectory \( q(t),p(t) \) is a solution of the Hamilton equations of motion:

\begin{eqnarray*}
\partial _{t}q=\partial _{p}H(q,p), &  & \\
\partial _{t}p=-\partial _{q}H(q,p). &  & 
\end{eqnarray*}

The trajectories \( q(t),p(t) \) evolve on the plane \( I\! \! R^{2} \), but
because of periodicity, they can be considered as trajectories on the torus
\( T_{qp}, \) obtained by identifying the opposite sides of the cell \( [0,1]\times [0,1]. \)

\subsection{Properties on trajectories}

Because of conservation of energy: \( d(H(q(t),p(t)))/dt=0 \), each trajectory
is included in an energy level \( \Sigma _{E}=\{(q,p)/H(q,p)=E\} \) of the
Morse function \( H \). Moreover, the velocity on such a trajectory is never
zero, except when there is a critical point of \( H \) where \( dH=0 \). 

We can easily obtain properties on the topology of \( \sum _{E} \):

\begin{enumerate}
\item For non critical value of \( E \), \( \Sigma _{E} \) is a closed one dimensional
curve, and each trajectory can be identified with a connected component of \( \Sigma _{E} \).
Therefore, each trajectory is a periodic orbit, an oriented closed curve on
the torus \( T_{qp} \). Its topology (homology) is characterized by two integers
\( (n_{1},n_{2})\epsilon Z^{2} \)which are the degrees in the \( q \) and
\( p \) directions of \( q(t),p(t) \) for \( t=0\rightarrow T \) where \( T \)
is the period. \( (n_{1},n_{2}) \) are relatively primes, except if \( n_{1}=0 \)
(or \( n_{2}=0 \)) in which case \( n_{2}=-1,0 \) or \( +1 \). A contractible
trajectory has homology type \( n_{1}=n_{2}=0 \) and is clockwise or anti-clockwise. 
\item A generic Hamiltonian has always non contractible closed trajectories with type
\( (n_{1},n_{2})\neq (0,0) \). In this case, every other closed trajectory
is one of the three types \( (n_{1},n_{2}),(-n_{1},-n_{2}) \) or \( (0,0) \),
and each of these three types exists. (Example: consider a perturbation of \( H(q,p)=\cos (n_{2}q-n_{1}p) \)). 
\item For non critical value of \( E \), each trajectory belongs locally to a family
parameterized by the energy \( E \) such that every trajectory has the same
type. (The critical trajectories are bifurcations between these families). 
\end{enumerate}
For a given generic Hamiltonian, by a good choice of the lattice generators
on the plane \( (q,p) \), it is possible to deal with trajectories of types
\( (0,0) \) \( (0,+1) \) and \( (0,-1) \) only. We will adopt this choice
in the sequel of this paper.

Now concerning the critical points and the critical trajectories :

\begin{enumerate}
\item There are three sorts of critical points: a local minimum, a saddle point or
a local maximum. From Euler formula (\cite{milnor-morse} p.29) we have \( \#(mins)+\#(maxs)-\#(saddles)=0 \)
(The Euler characteristic of the torus is \( 0 \)). 
\end{enumerate}

\subsection{Example}

We will give now an example in order to illustrate these statements. This example
will be used throughout the paper.

Consider the following Hamiltonian: 
\begin{eqnarray}
H(q,p)=H_{0}(q,p)+H_{1}(q,p), &  & \label{e:exemple_H} \\
H_{0}(q,p)=\cos (2\pi q)+0.1\cos (2\pi p), &  & \nonumber \label{math-nonumber} \\
H_{1}(q,p)=P\left[ \exp \left( -100(q-q_{0})^{2}-10(p-p_{0})^{2}\right) \right] , &  & \nonumber 
\end{eqnarray}
 where \( P \) is a functional operator which sums the function in each cell
and makes it periodic: 
\[
P\left[ f(q,p)\right] =\sum _{i,j}f(q+i,p+j).\]

\begin{figure}[h]
{\par\centering \resizebox*{0.5\columnwidth}{!}{\includegraphics{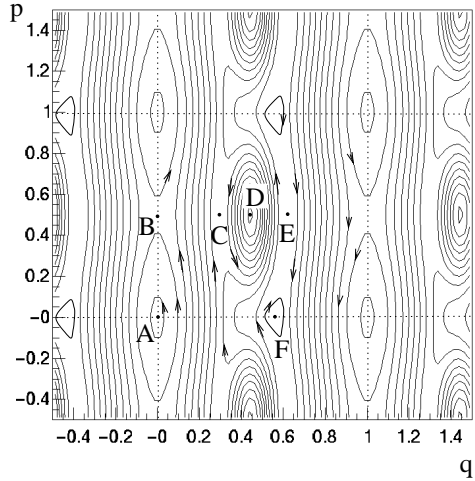}} \par}

\caption{\label{fig:levels}Trajectories and fixed points of Hamiltonian (\ref{e:exemple_H}).}
\end{figure}

The trajectories generated by \( H \) are shown on the figure \ref{fig:levels},
for \( q_{0}=0.45 \), \( p_{0}=0.45 \). There are one minimum F, two maxima
A,D, and three saddle points B,C,E.

\begin{figure}[h]
{\par\centering \resizebox*{0.5\columnwidth}{!}{\includegraphics{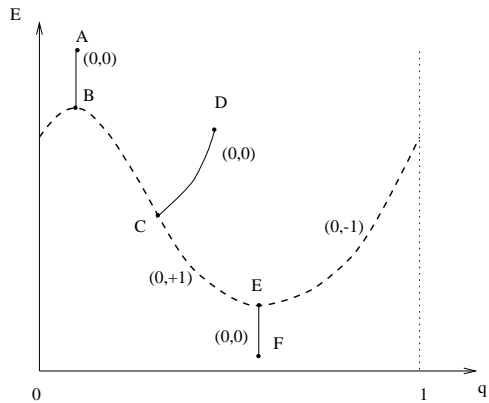}} \par}

\caption{\label{fig:bifurc}Reeb graph of \protect\( H\protect \): a schematic view
of the Trajectories types and bifurcations of Hamiltonian (\ref{e:exemple_H}).
Each point of the dashed line corresponds to a non contractible trajectory.
They form a one-dimensional family. Each point of a solid line corresponds to
a contractible trajectory.}
\end{figure}

The figure \ref{fig:bifurc} gives a schematic view of the families of trajectories,
with their bifurcations, and homology types. It is called a ``Reeb graph''
\( \mathcal{R} \) \cite{toulet1}. By definition, it is the topological quotient
space \( \mathcal{R}=T_{qp}/\mathcal{F} \) where \( \mathcal{F}=\{\Sigma _{E}\}_{E} \)
is the set of energy levels. The topology of this graph gives the relative location
of the trajectories (the \( q \) variable here is only to indicate that the
dashed line can be obtained by a section at constant \( p \) in figure \ref{fig:levels}).

The Reeb graph can be done for any given generic Hamiltonian. This figure will
always have a similar apparency, with one transverse (dashed) line (from \( q=0 \)
to \( q=1 \)) corresponding to the non contractible trajectories, and branches
from it (solid lines), corresponding to contractible trajectories. Critical
energies are at the extremities of these branches. The final value of the Chern
indices will be read more or less directly from this graph in section 7.

\section{The Quantum Hamiltonian}

In this section we introduce well known material that will be useful for the
sequel.

\subsection{The Hilbert space \protect\( \mathcal{H}_{P}\protect \) of the plane \label{s:plane}}

The Hilbert space corresponding to the classical motion in the plane phase space
\( T^{*}(I\! \! R)\cong I\! \! R^{2} \) is 
\[
\mathcal{H}_{P}=L^{2}(I\! \! R),\]
 (the index \( P \) will remind us that we deal with the plane phase-space). 

We choose a symmetric quantization procedure: to the classical Hamiltonian \( H(q,p) \)
we associate a self-adjoint operator \( \hat{H} \) in \( \mathcal{H}_{P} \)
by 
\[
\hat{H}=\sum _{n_{1},n_{2}\in Z^{2}}\frac{1}{2}c_{n_{1},n_{2}}\exp \left( i2\pi n_{1}\hat{q}\right) \exp \left( i2\pi n_{2}\hat{p}\right) +\textrm{hermitian conjugate}.\]

Where \( c_{n_{1}n_{2}} \) are the Fourier components of the function \( H(q,p) \):

\begin{equation}
\label{e:hc}
H(q,p)=\sum _{n_{1},n_{2}\in Z^{2}}c_{n_{1},n_{2}}\exp (i2\pi n_{1}q)\exp (i2\pi n_{2}p).
\end{equation}

Since \( H \) is a real valued function, the complex coefficients \( c_{n_{1}n_{2}} \)
must satisfy: 
\[
c_{n_{1},n_{2}}=\bar{c}_{-n_{1},-n_{2}}\in I\! \! \! \! C,\quad (n_{1},n_{2})\in Z\! \! \! Z^{2}.\]

We denote by \( \hat{T}_{Q} \){[}respectively \( \hat{T}_{P} \){]} the translation
operator by one period: \( \hat{T}_{Q} \) translates by \( Q=1 \) a wave function
\( \psi (x) \) and \( \hat{T}_{P} \) translates its Fourier transform \( \hat{\psi }(p) \)
by \( P=1 \): 
\begin{eqnarray}
\hat{T}_{Q}\psi (x) & = & \psi (x-1)\label{eq:defTQ} \\
\hat{T}_{P}\hat{\psi }(p) & = & \hat{\psi }(p-1).\label{eq:defTP} 
\end{eqnarray}
 We may rewrite equations (\ref{eq:defTQ}) and (\ref{eq:defTP}) as: 
\[
\hat{T}_{Q}=\exp (-i\hat{p}/\hbar ),\quad \hat{T}_{P}=\exp (i\hat{q}/\hbar ).\]

Quantum mechanically speaking, the periodicity Eq. (\ref{e:periodic}) reads:
\begin{equation}
\label{e:periodicity2}
[\hat{H},\hat{T}_{Q}]=[\hat{H},\hat{T}_{P}]=0.
\end{equation}

To continue, we now have to assume that 
\begin{equation}
\label{eq:commutativite}
[\hat{T}_{Q},\hat{T}_{P}]=0.
\end{equation}
 It is easy to prove that: 
\[
\hat{T}_{Q}\hat{T}_{P}=e^{-i/\hbar }\hat{T}_{P}\hat{T}_{Q},\]
 hence (\ref{eq:commutativite}) is fulfilled if and only if there exists an
integer \( N \) such that: 
\begin{equation}
\label{e:hyp}
N=\frac{1}{h}\in I\! \! N^{*}.
\end{equation}
with \( h=2\pi \hbar  \). This hypothesis (\ref{e:hyp}) can be regarded as
a geometric quantization condition, which states that there is an integer number
of Planck cells in the phase space. The semi-classical limit \( \hbar \rightarrow 0 \)
corresponds then to the limit \( N\rightarrow +\infty  \).

We now assume that hypothesis (\ref{e:hyp}) is fulfilled.

Note that hypothesis is not so restrictive, because any value of \( h \) can
be well approximated by a rational number \( h=m/N \). It is then easy to check
that \( [\hat{T}_{mQ},\hat{T}_{P}]=0 \). In this case we just have then to
consider the torus phase space \( [0,mQ[\times |0,P[ \) instead of \( [0,Q[\times |0,P[ \).

\subsection{The Hilbert space \protect\( \mathcal{H}_{T}(\theta _{1},\theta _{2})\protect \)
for the torus }

According to the commutation relations (\ref{e:periodicity2}) and (\ref{eq:commutativite}),
the Hilbert space \( L^{2}(I\! \! R) \) may be decomposed as a direct sum of
the eigen-spaces of the translation operators \( \hat{T}_{Q} \) and \( \hat{T}_{P} \):
\begin{eqnarray}
\mathcal{H}_{P}=L^{2}(I\! \! R) & = & \int \int \mathcal{H}_{T}(\theta _{1},\theta _{2})\, \, \frac{d\theta _{1}d\theta _{2}}{(2\pi )^{2}}\label{e:decomposition} \\
\mathcal{H}_{T}(\theta _{1},\theta _{2}) & = & \left\{ |\psi >\textrm{ such that }\left\{ \begin{array}{c}
\hat{T}_{Q}|\psi >=\exp (i\theta _{1})|\psi >\\
\hat{T}_{P}|\psi >=\exp (i\theta _{2})|\psi >
\end{array}\right. \right\} ,\nonumber 
\end{eqnarray}
 with \( (\theta _{1},\theta _{2})\in [0,2\pi [^{2} \) related to the periodicity
of the wave function under translations by an elementary cell. The space of
the parameters \( (\theta _{1},\theta _{2}) \) has also the topology of a torus,
and will be denoted by \( T_{\theta }. \)

The space \( \mathcal{H}_{T}(\theta _{1},\theta _{2}) \) is not a subspace
of \( L^{2}(I\! \! R) \), the space of physical states, it is a space of distributions
in the \( x \) representation. 

It is well known that \( H_{T}(\theta _{1},\theta _{2}) \) is finite dimensional
\cite{iqhe2}. To see that, let \( |\psi >\in \mathcal{H}_{T}(\theta _{1},\theta _{2}) \).
The Fourier transform of \( \psi (x) \) is \( \theta _{2} \)-Floquet-periodic
of period \( P \), so \( \psi (x) \) is discrete, it is a sum of Dirac distributions
supported at points distant from \( h=1/N \) from each other. Moreover, \( \psi (x) \)
is \( \theta _{1} \)-periodic, hence \( \psi  \) is characterized by the \( N \)
coefficients at the \( N \) Dirac distributions supporting points in the interval
\( q\in [0,1[ \).

Explicitly, a basis of \( \mathcal{H}_{T}(\vec{\theta }) \) is given by \( N \)
distribution denoted \( |j,\vec{\theta }> \): 
\begin{eqnarray*}
|j,\vec{\theta }>\equiv \psi _{j,\vec{\theta }}(x)=\frac{1}{\sqrt{N}}\sum _{n_{1}\in Z\! \! \! Z}\exp \left( -in_{1}\theta _{1}\right) \delta (x-q_{j}-n_{1}),\quad j=1,\ldots ,N &  & \\
\textrm{with }q_{j}=\frac{1}{N}\left( j+\frac{\theta _{2}}{2\pi }\right) . &  & 
\end{eqnarray*}

Eventually we get: 
\[
\dim _{I\! \! \! \! C}\, \, \mathcal{H}_{T}(\theta _{1},\theta _{2})=N.\]

Because of Eq.(\ref{e:periodicity2}) the Hamiltonian \( \hat{H} \) is block-diagonal
with respect to the decomposition Eq.(\ref{e:decomposition}). The operator
\( \hat{H} \) acts on \( \mathcal{H}_{T}(\theta _{1},\theta _{2}) \) as a
\( N\times N \) hermitian matrix, its spectrum \( \sigma (\theta _{1},\theta _{2}) \)
is made of \( N \) eigenvalues and \( N \) corresponding eigen-functions:
\begin{equation}
\label{e:spectre}
\hat{H}|\varphi _{n}(\theta _{1},\theta _{2})>=E_{n}(\theta _{1},\theta _{2})|\varphi _{n}(\theta _{1},\theta _{2})>,\quad n=1,\ldots ,N.
\end{equation}

For a given level \( n, \) assuming that \( E_{n}(\theta _{1},\theta _{2}) \)
is never degenerate \( \forall \theta  \), the energy level \( E_{n}(\theta _{1},\theta _{2}) \)
forms a band as \( (\theta _{1},\theta _{2})\in T_{\theta } \) are varying,
and the eigenvectors \( |\psi _{n}(\theta _{1},\theta _{2})> \) form a \( 2D \)
surface in the quantum states space. But for a fixed \( (\theta _{1},\theta _{2}), \)
and any \( \lambda \in I\! \! \! \! C \), \( \lambda |\psi _{n}(\theta _{1},\theta _{2})> \)
is also an eigenvector. So for a fixed level \( n \), the family of eigenvectors
form a complex-line-bundle (of fiber \( \cong I\! \! \! \! C\ni \lambda  \)
) in the projective space of the bundle \( \mathcal{H}_{T}\rightarrow T_{\theta } \).
The topology of this line bundle is characterized by an integer \( C_{n}\in Z\! \! \! Z \),
called the Chern index (\cite{harris1}, chap 1, p.139).

The definition of this topological Chern index is presented in section \ref{s:def}.

\subsection{The Hilbert space \protect\( \mathcal{H}_{C}(\theta _{2})\protect \) for the
cylinder }

In the sequel it will be useful to consider an intermediate decomposition of
\( \mathcal{H}_{P} \) in terms of states periodic in momentum only: 
\begin{eqnarray*}
\mathcal{H}_{P}=L^{2}(I\! \! R)=\int \mathcal{H}_{C}(\theta _{2})\frac{d\theta _{2}}{2\pi } &  & \\
\mathcal{H}_{C}(\theta _{2})=\left\{ |\psi >\textrm{ such that }\hat{T}_{P}|\psi >=\exp (i\theta _{2})|\psi >\right\} . &  & 
\end{eqnarray*}
 Each state in \( \mathcal{H}_{C}(\theta _{2}) \) is a sum of Dirac distributions
supported at points \( q_{n}=\frac{1}{N}\left( n+\frac{\theta _{2}}{2\pi }\right)  \)
with \( n\in Z\! \! \! Z \) . The weights belongs to \( l^{2}(Z\! \! \! Z) \).

Conversely: 
\[
\mathcal{H}_{C}(\theta _{2})=\int \mathcal{H}_{T}(\theta _{1},\theta _{2})d\theta _{1}.\]

For each \( \theta _{2} \), the Hilbert space \( \mathcal{H}_{C}(\theta _{2}) \)is
associated with the cylinder phase space denoted \( C_{qp} \), where \( (q,p)\in I\! \! R^{2} \)
is identified with \( (q,p+1) \).

From a state \( |\psi _{P}>\in \mathcal{H}_{P}=L^{2}(I\! \! R) \) of the plane,
we can construct a state \( |\psi _{C}>\in \mathcal{H}_{C}(\theta _{2}) \)
of the cylinder by the operation: 
\begin{equation}
\label{e:cyl}
|\psi _{C}(\theta _{2})>=P_{_{\theta _{2}}}|\psi _{P}>=\sum _{n=-\infty }^{+\infty }\exp (-in\theta _{2})T_{P}^{n}|\psi _{P}>.
\end{equation}

From a state \( |\psi _{C}>\in \mathcal{H}_{C}(\theta _{2}) \) of the cylinder,
we can construct a state \( |\psi _{T}(\theta _{1},\theta _{2})>\in \mathcal{H}_{T}(\theta _{1},\theta _{2}) \)
of the torus by the operation: 
\begin{equation}
\label{e:torus}
|\psi _{T}(\theta _{1},\theta _{2})>=P_{_{\theta _{1}}}|\psi _{C}(\theta _{2})>=\sum _{n=-\infty }^{+\infty }\exp (-in\theta _{1})T_{Q}^{n}|\psi _{C}(\theta _{2})>.
\end{equation}

\section{The quasi modes }

\label{s:quasi-modes}

\subsection{Construction of a quasi-mode}

The W.K.B. method allows the construction of a quasi-modes for integrable dynamics.
See \cite{arnold1}, Lazutkin \cite{lazutkin1} p. 235.

First, we recall the definition and general properties of quasi-modes.

\begin{defn}
If \( \Gamma  \) is a given closed trajectory with energy \( E \), a quasi-mode
with error \( \epsilon  \), is a quantum state \( |\tilde{\psi }> \) localized
near \( \Gamma  \), which satisfies:
\[
\left\Vert \hat{H}|\tilde{\psi }>-\tilde{E}|\tilde{\psi }>\right\Vert =\epsilon =o(h^{\infty }),\]
 with \( h=1/N \) going to zero, with \( (E-\tilde{E})=o(h) \). The trajectory
\( \Gamma  \) is called the support of \( |\tilde{\psi }> \) and  noted 
\[
\Gamma =\textrm{Supp}(|\tilde{\psi }>).\]

\end{defn}
If we want to improve this error \( \epsilon  \) (and have an error exponentially
small with respect to \( h \)), we have to take into account tunneling effect
between \( \Gamma  \), and \( \Gamma ' \) the ``closest'' trajectory \( \Gamma ' \)
which has the same energy \( E \). (\cite{HS},\cite{harper8},\cite{harper4}).
We will come back on this point.

\begin{prop}
\textbf{\label{prop:quasi-mode1}}\cite{lazutkin1}

The interval \( [\tilde{E}-\epsilon ,\tilde{E}+\epsilon ] \) contains at least
one eigenvalue of \( \hat{H} \). 

Let \( \Delta _{\alpha }=[\tilde{E}-\alpha ,\tilde{E}+\alpha ] \). If \( \Delta _{\alpha } \)
contains only one eigen-value \( E^{*} \), with eigen-vector \( |\varphi > \),
then (for normalized vectors)
\begin{equation}
\label{e:inegalite}
\exists \beta \in R,\qquad \left\Vert |\varphi >-e^{i\beta }|\tilde{\psi }>\right\Vert \leq \frac{\epsilon }{\alpha },
\end{equation}
consequently, \( <\varphi |\tilde{\psi }>\neq 0 \) as soon as \( \epsilon \leq \alpha \sqrt{2} \)
(for our purpose, we will say in that case, that \( |\tilde{\psi }> \) is a
``good approximation'' of \( |\varphi > \). This condition will be sufficient
to construct a bundle of quasi-modes with the same topology, same Chern index).
\end{prop}
Let us now give properties of quasi-modes specific to our problem. The explicit
construction of the quasi-modes can be found in \cite{voros1}\cite{wkb2}\cite{wkb3}.

\subsubsection*{Contractible trajectories of type \protect\( (0,0)\protect \)}

\begin{figure}[h]
{\par\centering \resizebox*{0.5\columnwidth}{!}{\includegraphics{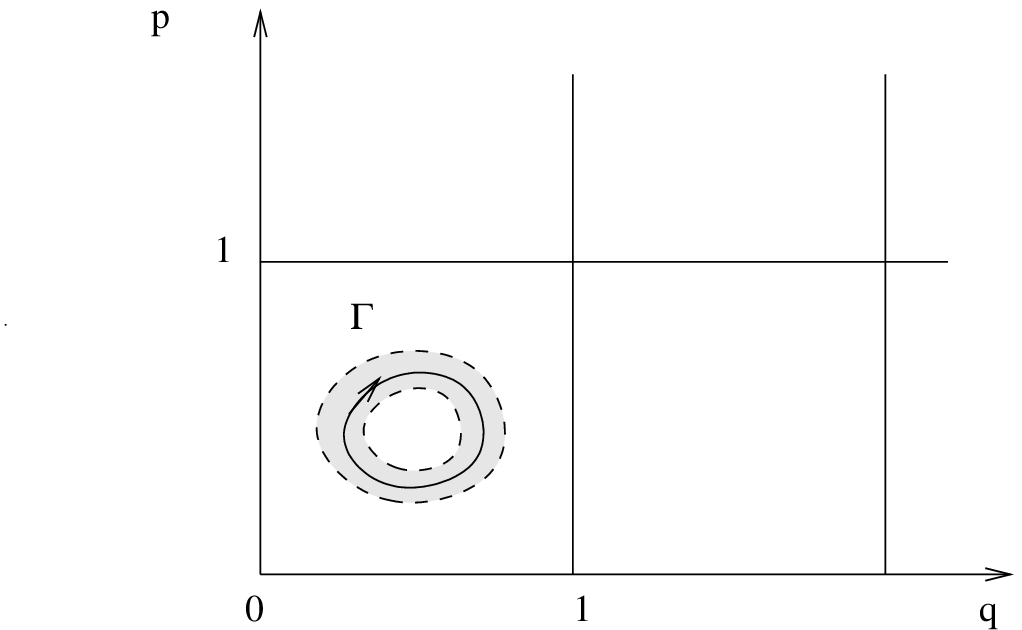}} \par}

\caption{\label{fig:quasi_mode00} Quasi-mode localized on a trajectory \protect\( \Gamma \protect \)
of type \protect\( (0,0)\protect \) in cell \protect\( (0,0)\protect \).}
\end{figure}

For a non critical contractible trajectory \( \Gamma  \) of type \( (0,0) \),
we first construct a quasi-mode on the plane, \( |\tilde{\psi }_{P}>\in \mathcal{H}_{P} \).
From the usual W.K.B construction, this quasi mode is localized on a single
image of \( \Gamma  \), for example in the cell \( (0,0) \). See figure \ref{fig:quasi_mode00}.

The energy \( \tilde{E} \) of this quasi-mode is given by the usual E.B.K.
condition: 
\begin{eqnarray*}
S(\tilde{E})=(k+1/2)h+o(h) & ,\quad k\in Z\! \! \! Z, & \\
 & 
\end{eqnarray*}
 with \( S(\tilde{E}) \) the surface inclosed in the trajectory \( \Gamma  \).

A quasi-mode \( |\tilde{\psi }_{C}(\theta _{2})> \) on the cylinder can be
deduced by eq.(\ref{e:cyl}). A quasi-mode \( |\tilde{\psi }_{T}(\theta _{1},\theta _{2})> \)
on the torus can be constructed by eq.(\ref{e:torus}). Note that the energy
of these quasi-modes \( \tilde{E} \) doesn't depend on \( (\theta _{1},\theta _{2}) \).

\subsubsection*{Non contractible trajectories of type \protect\( (0,\pm 1)\protect \)}

\begin{figure}[h]
{\par\centering \resizebox*{0.5\columnwidth}{!}{\includegraphics{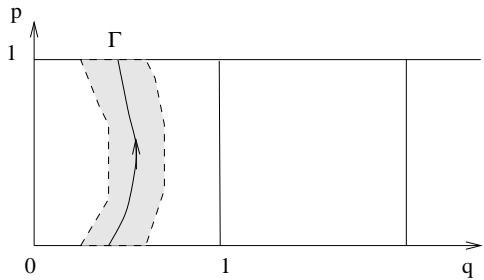}} \par}

\caption{\label{fig:quasi_mode01}Quasi-mode localized on a trajectory \protect\( \Gamma \protect \)
of type \protect\( (0,+1)\protect \), in cell \protect\( (0)\protect \) on
the cylinder.}
\end{figure}

If \( \Gamma  \) is a non critical trajectory with type \( \vec{n}=(0,\pm 1) \),
\( \Gamma  \) is closed on the cylinder \( C_{qp} \) and a quasi-mode \( |\tilde{\psi }_{C}(\theta _{2})>\in \mathcal{H}_{C}(\theta _{2}) \)
can be constructed for every \( \theta _{2} \).

\( |\tilde{\psi }_{C}(\theta _{2})> \) is localized on \( \Gamma  \), for
example in cell \( 0 \), see figure \ref{fig:quasi_mode01}.

The energy \( \tilde{E} \) of this quasi-mode is given by the condition: 
\begin{equation}
\label{e:quantif}
S(\tilde{E})=(k-\theta _{2}/2\pi )h+o(h),\quad k\in Z\! \! \! Z,
\end{equation}

where \( S(\tilde{E}) \) is the surface included in the unit cell, on the right
of the trajectory \( \Gamma  \) (here oriented by increasing \( p \)). We
justify this condition in the next paragraph.

Note that the energy \( \tilde{E}(\theta _{2}) \) as well as the support \( \Gamma (\theta _{2}) \)
depend continuously on \( \theta _{2} \).

A quasi mode \( |\tilde{\psi }_{T}(\theta _{1},\theta _{2})> \) on the torus
can be deduced from \( |\tilde{\psi }_{C}(\theta _{2})> \)by eq.(\ref{e:torus}).

For contractible and non contractible trajectories, an other quasi-mode with
the same energy can be constructed in the cell \( n \) by
\begin{equation}
\label{e:trans}
|\tilde{\psi }_{C,n}(\theta _{2})>=T_{Q}^{n}|\tilde{\psi }_{C}(\theta _{2})>.
\end{equation}

From the W.K.B. construction, the error \( \epsilon  \) of these quasi-modes
is of order \( o(h^{\infty }) \). The error \( \epsilon  \) is due to tunneling
interaction with ``the nearest quasi-mode of nearby energy'' (see the next
section).

\subsubsection*{Critical trajectories}

If \( E_{0} \) is a critical energy, quasi modes of energy near \( E_{0} \)
(in the interval \( [E_{0}\pm O(h)] \)) and localized near the critical trajectory,
can be constructed. See \cite{CPI}\cite{CPII}\cite{CPIII}.

\subsubsection*{Justification of quantization rule eq.\ref{e:quantif}}

When constructing a WKB quasi-mode on a closed trajectory, the quantization
rule comes from the phase matching. The phase of the Fourier transform of the
quasi-mode evolves like \( (-\int qdp) \) along the trajectory \( q(t),p(t) \).
For a trajectory \( \Gamma  \) of type \( (0,\pm 1) \), the phase accumulated
over a period is then \( \varphi =S/\hbar =-\int _{\Gamma }qdp/\hbar  \), where
\( S \) is the surface on the right of the trajectory \( \Gamma  \) (oriented
by increasing value of \( p \)). The periodicity condition (\ref{e:decomposition})
requires that \( \varphi =\theta _{2}\quad [2\pi ] \). This gives the result
(\ref{e:quantif}).

\subsection{The semi-classical spectrum without tunneling corrections}

We now investigate the global properties of the quasi-modes, with respect to
\( \theta _{2} \).

Consider fixed values of \( (\theta _{1},\theta _{2})\in T_{\theta } \). A
W.K.B. construction as previously described, gives us \( N \) quasi-modes \( |\tilde{\psi }_{T}(\theta _{1},\theta _{2})>_{n} \)
with energies \( \tilde{E}_{n}(\theta _{1},\theta _{2}) \) over the whole energy
range, with \( n=1\rightarrow N \). This is a semi-classical approximation
\( \sigma _{sc}(\theta _{1},\theta _{2}) \) of the spectrum \( \sigma (\theta _{1},\theta _{2}) \)
defined by eq.(\ref{e:spectre}). In fact the construction was made on \( \mathcal{H}_{C}(\theta _{2}) \)
and does not depend on \( \theta _{1} \), so we will note it \( \sigma _{sc}(\theta _{2}) \).
(For contractible trajectories, it does not even depend on \( \theta _{2} \)).

From the above construction, each quasi-mode \( |\tilde{\psi }_{C}(\theta _{2})> \)
depends continuously on \( \theta _{2} \). So each quasi-mode belongs to a
function \( \Psi :\theta _{2}\rightarrow |\tilde{\psi }_{C}(\theta _{2})>\in \mathcal{H}_{C}(\theta _{2}) \)
which is periodic with respect to \( \theta _{2} \) (up to a phase), with a
period \( \Theta _{2}=2\pi m \), with \( m\in N \). (Otherwise, if the function
\( \Psi  \) were not periodic, the spectrum \( \sigma _{sc}(\theta _{2}) \)
would be infinite for a fixed value of \( \theta _{2} \)). The corresponding
energy \( \tilde{E}(\theta _{2}) \) and the support \( \Gamma (\theta _{2}) \)
are therefore also periodic with the same periods.

If the trajectory \( \Gamma  \) is contractible (type \( (0,0) \)), the energy
\( \tilde{E} \) and the support \( \Gamma  \) are fixed. They do not depend
on \( \theta _{2} \), and the period is \( \Theta _{2}=2\pi  \).

\begin{figure}[h]
{\par\centering \resizebox*{0.5\columnwidth}{!}{\includegraphics{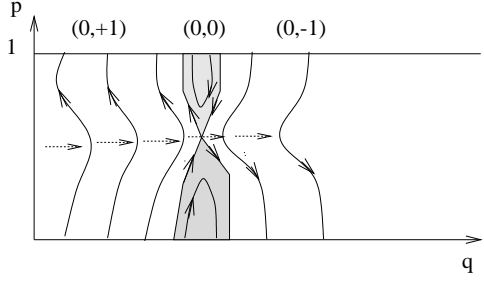}} \par}

\caption{\label{fig:traj_non_cont}The dashed arrows indicate the motion of the support
of a quasi mode, \protect\( \theta _{2}\rightarrow \Gamma (\theta _{2})\protect \),
over all non contractible trajectories, as \protect\( \theta _{2}\protect \)
increases.}
\end{figure}

For a non contractible and non critical trajectory (type \( (0,\pm 1) \)),
eq.(\ref{e:quantif}) tells that the surface \( \theta _{2}\rightarrow S(\theta _{2}) \)
on the right of \( \Gamma  \) is a strictly decreasing function because \( dS/d\theta _{2}=-h/2\pi  \).
Then \( \theta _{2}\rightarrow \Gamma (\theta _{2}) \) moves from the left
to the right on the set of non contractible trajectories on the cylinder \( C_{qp} \).
This function reaches all the non contractible trajectories, and jumps over
a connected component of contractible trajectories, via a critical point . See
figure \ref{fig:traj_non_cont}.

\begin{figure}[h]
{\par\centering \resizebox*{0.6\columnwidth}{!}{\includegraphics{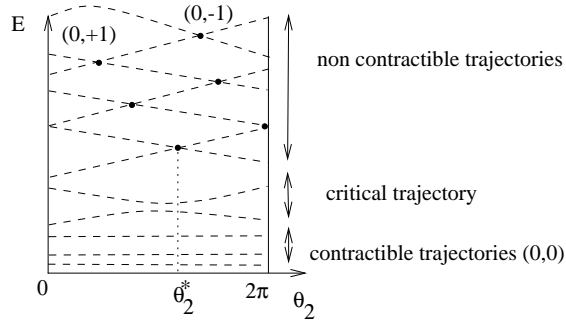}} \par}

\caption{\label{fig:sc_energy}The energy of the quasi-modes as a function of \protect\( \theta _{2}\protect \).}
\end{figure}

The surface \( S(\theta _{2}) \) is always a decreasing function, but the corresponding
energy \( E(\theta _{2}) \) decreases for trajectories of type \( (0,+1) \)
and increases for type \( (0,-1) \), with the behavior shown on figure \ref{fig:bifurc},
if we plot \( E(\theta _{2}) \) for \( \theta _{2}=0\rightarrow 2\pi m \).
Figure \ref{fig:sc_energy} shows this function \( E(\theta _{2}) \) but folded
in the interval \( \theta _{2}=0\rightarrow 2\pi  \).

So we obtain:

\begin{prop}
\textbf{\label{p:prop_cycles}}

The global spectrum \( \theta _{2}\rightarrow \sigma _{sc}(\theta _{2}) \)
is the union of \textbf{one single} function \( \Psi _{nc}:\, \theta _{2}\rightarrow |\tilde{\psi }_{nc}(\theta _{2})>\in \mathcal{H}_{C}(\theta _{2}) \),
and many others \( (\Psi _{c})_{i=1,..,N_{c}}:\, \theta _{2}\rightarrow |\tilde{\psi }_{c}(\theta _{2})>\in \mathcal{H}_{C}(\theta _{2}) \)
such that

The support \( \Gamma _{nc} \) of \( \Psi _{nc} \) is on the non contractible
trajectories. The period of \( \Psi _{nc} \) \emph{}is \( \Theta _{2}=2\pi .N_{nc} \),
such that \( N_{nc}.h\sim S_{nc} \) the surface occupied by the non-contractible
trajectories. In phase space, this support \emph{\( \theta _{2}\rightarrow \Gamma _{nc}(\theta _{2}) \)}
is a closed cycle of homotopy 1 on the cylinder \( C_{qp} \). 

The support \emph{\( \Gamma _{c} \)} of each function \( (\Psi _{c})_{i} \)
is on a fixed contractible trajectory. Its period is \( \Theta _{2}=2\pi  \).
The number \( N_{c} \) of these functions is such that \( N_{c}.h\sim S_{c} \)
the surface occupied by the contractible trajectories. Note that \( N=N_{c}+N_{nc} \)
and \( 1=S_{nc}+S_{c} \).
\end{prop}
We now discuss the error of the spectrum \( \sigma _{sc}(\theta _{2}) \).

The error of each quasi-mode is \( \epsilon \sim o(h^{\infty }) \). There are
crossings in the spectrum \( \sigma _{sc}(\theta _{2}) \) for discrete values
\( \theta _{2}^{*} \), (caused by the non contractible function when \( \tilde{E}_{nc}(\theta _{2})=\tilde{E}_{nc}(\theta _{2}^{'}) \)
or \( \tilde{E}_{nc}(\theta _{2})=\tilde{E}_{c}(\theta _{2}^{'}) \)), see figure
\ref{fig:sc_energy}. For fixed \( \theta _{2} \), away from a small neighborhood
of every crossing value \( \theta _{2}^{*} \), the spectrum \( \sigma _{sc}(\theta _{2}) \)
has \( N \) discrete eigenvalues and each eigen-value is isolated from the
overs by an interval of length \( \alpha \gg \epsilon  \). We deduce from properties
\ref{prop:quasi-mode1} that the corresponding quasi-modes \( |\tilde{\psi }_{n}(\theta _{2})> \)
(in fact their image in \( \mathcal{H}_{T} \) by the mapping \( P_{\theta _{2}} \)),
are a good approximation of the actual eigenvectors \( |\varphi _{n}(\theta _{1},\theta _{2})> \),
for any \( \theta _{1} \). 

Let us remark that this semi-classical spectrum \( \sigma _{sc}(\theta _{2}) \)
does not depend on \( \theta _{1} \) and is thus infinitely degenerate. This
correspond to the invariance by the translation (\ref{e:trans}). This is not
the case in the actual spectrum \( \sigma (\theta _{1},\theta _{2}) \) because
of tunneling effect.

Figure \ref{fig:bandes} shows the numerical spectrum of our example \( H(q,p) \),
eq.(\ref{e:exemple_H}). From this, we deduce what the semi-classical spectrum
should look like: fig (\ref{fig:bandes_sc}). Compare this with the semi-classical
spectrum figure \ref{fig:sc_energy}.

\( \sigma _{sc}(\theta _{2}) \) is not a good approximation of the actual spectrum
near the crossing values \( \theta _{2}^{*} \). To improve that, in the next
section, we will take into account the tunneling effect at each crossing of
energy. It causes a splitting of levels and we will recover \( N \) well defined
bands of energy.

\subsection{The semi-classical spectrum with tunneling corrections}

Consider a neighborhood \( U \) of a value \( \theta _{2}^{*} \), where there
is a crossing in the spectrum \( \sigma _{sc}(\theta _{2}) \) between two energies:
\( \tilde{E}_{1}(\theta _{2}^{*})=\tilde{E}_{2}(\theta _{2}^{*}) \). See figure
\ref{fig:sc_energy}. As said before, this crossing is caused two non contractible
trajectories when \( \tilde{E}_{nc}(\theta _{2})=\tilde{E}_{nc}(\theta _{2}^{'}) \)
or by one non contractible trajectory and one contractible one: \( \tilde{E}_{nc}(\theta _{2})=\tilde{E}_{c} \).
The crossing due to two contractible trajectories \( \tilde{E}_{c1}=\tilde{E}_{c2} \)
is not considered here, because it does not occur for a generic Hamiltonian
\( H \). See \cite{fred1}\cite{fred2} for results treating this non generic
case. From eq.(\ref{e:trans}), these energies correspond to two families of
quasi-modes: \( |\tilde{\psi }1_{C,n_{1}}> \), \( |\tilde{\psi }2_{C,n_{2}}>\in \mathcal{H}_{C}(\theta _{2}) \)
where \( n_{1}\in Z\! \! \! Z \) , \( n_{2}\in Z\! \! \! Z \). We suppose
that these quasi-modes are localized respectively on two trajectories \( \Gamma _{1} \)
and \( \Gamma _{2} \), in cells \( n_{1} \) and \( n_{2} \).

\begin{figure}[h]
{\par\centering \resizebox*{0.5\columnwidth}{!}{\includegraphics{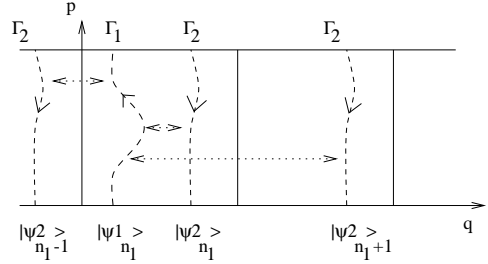}} \par}

\caption{\label{fig:tunnel}Tunneling effect between quasi-mode \protect\( |\psi 1_{n1}>\protect \)
on trajectory \protect\( \Gamma _{1}\protect \), in cell \protect\( n_{1}\protect \),
and quasi-modes \protect\( |\psi 2_{n2}>\protect \). We keep only the dominant
tunneling interaction, here due to \protect\( |\psi 2_{n1}>\protect \).}
\end{figure}

To improve the error \( \epsilon  \) of the quasi-mode say \( |\tilde{\psi }1_{C,n_{1}}> \),
we have to take into account the tunneling interactions with other quasi-modes
which have nearby energy \( |\tilde{\psi }j_{C,m}> \)with \( j=1,2 \) and
\( m\in Z\! \! \! Z \) . See figure \ref{fig:tunnel}. This tunneling interaction
comes from non vanishing terms \( A_{j,m}=<\tilde{\psi }1_{C,n_{1}}|\hat{H}|\tilde{\psi }j_{C,m}> \).
We will treat the tunneling effect at the leading order, by keeping only the
dominant term. Consider first the terms \( A_{2,n_{2}} \) with \( j=2 \),
and \( n_{2}\in Z\! \! \! Z \). We keep the \( n_{2} \) term with greatest
modulus: \( |A_{2,n_{2}}| \). This term is \( o(h^{\infty }) \) and describe
the tunneling interaction between \( |\tilde{\psi }1_{C,n_{1}}> \)and the ``nearest''
quasi-mode \( |\tilde{\psi }2_{C,n_{2}}> \). See figure \ref{fig:tunnel}.
It is clear that the nearest quasi-mode is \( n_{2}=n_{1} \) or \( n_{2}=n_{1}\pm 1 \),
because other values of \( n_{2} \) are in more distant cells. The other terms
are \( o(h^{\infty }) \) with respect to leading order term. Consider now the
terms \( A_{1,m} \) with \( j=1 \), and \( m\in Z\! \! \! Z \). The dominant
term is for \( m=n_{1} \): \( A_{1,n_{1}}=<\tilde{\psi }1_{C,n_{1}}|\hat{H}|\tilde{\psi }1_{C,n_{1}}>=\tilde{E}_{1}(\theta _{2}) \).
This is the energy of the quasi-mode. The others terms, which describe tunneling
interaction between \( |\tilde{\psi }1_{C,n_{1}}> \) and its images \( |\tilde{\psi }1_{C,m}> \)
in other cells, are \( o(h^{\infty }) \) with respect to \( |A_{2,n_{2}}| \)
and so negligible. 

We have obtained that the dominant correction to the quasi-modes \( |\tilde{\psi }1_{C,n_{1}}> \)
is described by the \( 2\times 2 \) tunneling interaction matrix:
\[
A(\theta _{2})=\left( <\tilde{\psi }i_{C,n_{1}}|\hat{H}|\tilde{\psi }j_{C,n_{2}}>\right) _{i,j}=\left( \begin{array}{cc}
\tilde{E}_{1}(\theta _{2}) & A_{2,n_{2}}(\theta _{2})\\
\overline{A_{2,n_{2}}}(\theta _{2}) & \tilde{E}_{2}(\theta _{2})
\end{array}\right) ,\]
where \( |\tilde{\psi }2_{C,n_{2}}> \) is the ``nearest'' quasi-mode to \( |\tilde{\psi }1_{C,n_{1}}> \),
\( n_{2}=n_{1} \) or \( n_{2}=n_{1}\pm 1 \), and \( |A_{2,n_{2}}| \) is \( o(h^{\infty }) \).

By diagonalizing the matrix \( A \), we obtain two quasi-modes \( |\tilde{\psi }_{C+}(\theta _{2})> \)
and \( |\tilde{\psi }_{C-}(\theta _{2})> \)\( \in \mathcal{H}_{C}(\theta _{2}) \)
expressed as a linear superposition of \( |\tilde{\psi }1_{C,n_{1}}> \)and
\( |\tilde{\psi }2_{C,n_{2}}> \), and two energies \( E_{+}(\theta _{2})>E_{-}(\theta _{2}) \)
with a gap \( |\tilde{E}_{+}(\theta _{2})-\tilde{E}_{-}(\theta _{2})|\geq |A_{2,n_{2}}| \)
. But the error of the quasi-modes is \( o(h^{\infty }) \) with respect to
this gap. From property (\ref{prop:quasi-mode1}), we deduce that these quasi-modes
\( |\tilde{\psi }_{C\pm }(\theta _{2})> \) (their image in \( \mathcal{H}_{T} \)
actually) are a good approximation of the eigenvectors \( |\varphi _{i}(\theta _{1},\theta _{2})> \),
for any \( \theta _{1} \). 

Now if we match these results with the outside of the neighborhood \( U \)
of \( \theta _{2}^{*} \), we can obtain a continuous dependence of \( |\tilde{\psi }_{C}(\theta _{2})> \)with
respect to \( \theta _{2}\in I\! \! R \). 

\begin{prop}
\label{prop:chern_homotopy}We have obtained quasi-modes \( |\tilde{\psi }_{C}(\theta _{2})> \)
on the cylinder, which are a good approximation of the spectrum eq.(\ref{e:spectre})
for any \( \theta _{1},\theta _{2} \) in the sense that :\emph{
\begin{equation}
\label{e:sc-bands}
<\varphi _{n}(\theta _{1},\theta _{2})|\tilde{\psi }_{T}(\theta _{1},\theta _{2})>\neq 0\textrm{ for every }(\theta _{1},\theta _{2}).
\end{equation}
}

We note this approximate spectrum by \( \sigma _{sc,t}(\theta _{2}) \) (the
energies do not depend on \( \theta _{1} \)). 

There is no more crossing in the energy levels, so we have \( N \) well defined
(semi-classical) bands and corresponding quasi-modes \( |\tilde{\psi }_{C,n}(\theta _{2})> \),
\( n=1,..,N \).

At each avoided crossing, we have obtained that the support of a quasi-mode
jumps from the trajectory \( \Gamma _{1} \) in cell \( n_{1} \) (resp. \( \Gamma _{2} \)
in \( n_{2} \)) to the closest trajectory \( \Gamma _{2} \) in cell \( n_{2}=n_{1} \)
or \( n_{2}=n_{1}\pm 1 \) (respect. \( \Gamma _{1} \) in \( n_{1} \) ) as
in figure \ref{fig:tunnel}. 

More globally, for each band \( n \), we derived an function for the support
of the quasi-mode \( |\tilde{\psi }_{C,n}(\theta _{2})> \)on the cylinder \( C_{qp} \):
\begin{equation}
\label{e:support_n}
S_{Cn}:\quad \theta _{2}\in R\rightarrow \textrm{Supp}\left( |\psi _{C,n}(\theta _{2})>\right) \quad \subset C_{qp}.
\end{equation}

The image \( S_{T,n} \) of \( S_{C,n} \) on the torus is periodic for \( \theta _{2}=0\rightarrow 2\pi  \),
and \( S_{C,n} \) is a lifting of \( S_{T,n} \) on the cylinder. With \( S_{C,n} \)
is associated a homotopic number \( I_{n}\in Z\! \! \! Z \). Precisely, \( I_{n} \)
is given by: 
\begin{equation}
\label{e:defI}
T_{Q}^{I_{n}}\left[ \textrm{Supp}\left( |\psi _{C,n}(0)>\right) \right] =\left[ \textrm{Supp}\left( |\psi _{C,n}(2\pi )>\right) \right] .
\end{equation}

\( I_{n} \) is simply the result of the jumps of the quasi-mode from cell \( n_{1} \)
to cell \( n_{1}+I_{n} \) as \( \theta _{2}=0\rightarrow 2\pi  \).
\end{prop}

\section{Chern indices}

\subsection{Definition \label{s:def}}

The eigen-functions \( |\varphi _{n}(\theta _{1},\theta _{2})> \) of the band
\( n \), eq.(\ref{e:spectre}), form a complex line fiber bundle over the torus
\( (\theta _{1},\theta _{2})\in T_{\theta } \). On a contractible subset of
\( U\subset T_{\theta } \) we can choose the states \( |\varphi _{n}(\theta _{1},\theta _{2})> \)
such that they are normalized and that they form a (continuous) section over
\( U \). Note that this is not possible over the whole torus \( T_{\theta } \)
except if the bundle is trivial.

The global topology of this complex line bundle is characterized by its Chern
index (\cite{harris1} p. 139, \cite{avron1}).

Because of the natural Hilbert scalar product on \( L^{2}(I\! \! R)=\int \int \mathcal{H}_{T}(\theta _{1},\theta _{2})\, \, d\theta _{1}d\theta _{2} \),
which induces the Berry (or Chern) connection (\cite{berry1}, \cite{chern6}),
this topological number is explicitly given by the integral of the Berry (or
Chern) curvature (\cite{iqhe2},\cite{harris1} p. 141):

\begin{equation}
\label{e:curvature}
C_{n}=\frac{i}{2\pi }\int _{T_{\theta }}\left( <\partial _{\theta _{1}}\varphi _{n}|\partial _{\theta _{2}}\varphi _{n}>-<\partial _{\theta _{2}}\varphi _{n}|\partial _{\theta _{1}}\varphi _{n}>\right) d\theta _{1}d\theta _{2},
\end{equation}
 obtained by sum over local open subsets \( U_{i} \) which covered \( T_{\theta } \),
and sections chosen in each of them.

This expression has been used intensively for our numerical calculations.

Moreover, it can be shown (see e.g. \cite{iqhe1},\cite{chern3}) that: 
\begin{equation}
\label{eq:sigmacn}
\sum _{n=1}^{N}C_{n}=1.
\end{equation}

There is an alternative expression for \( C_{n} \) more suitable for our analytical
calculations, given in eq. (\ref{e:chern2}), (see \cite{harris1} p.141). It
is based on the motion of the zeros of the Bargmann representation of the states
\( |\varphi _{n}(\theta _{1},\theta _{2})> \).

\subsection{Chern indices of the semi-classical bands}

Our aim is to express the index \( C_{n} \) from the description of the classical
dynamics.

Because of property (\ref{e:sc-bands}), we deduce from theorem (\ref{th:homotopy})
page \pageref{th:homotopy},that the band of the semi-classical spectrum \( \sigma _{sc,t} \)
have the same topology (and same Chern indices) as the actual energy bands.

Our work consists now in computing the index \( C_{n} \) of the semi-classical
band \( n \).

From theorem (\ref{th:cylindre}) page \pageref{th:cylindre}, the result is
simply that: 

\begin{equation}
\label{e:result}
C_{n}=I_{n},
\end{equation}

where \( I_{n} \) defined by eq. (\ref{e:defI}), is the homotopic number characterizing
the path followed by the support of the quasi-mode on the cylinder, when \( \theta _{2} \)
is varying from \( 0 \) to \( 2\pi  \).

\subsection{Global analysis of the Chern indices}

In this section we calculate the sum of Chern indices for consecutive bands,
in order to recover the result eq.(\ref{eq:sigmacn}) within our semi-classical
approach.

For each band \( n=1\rightarrow N \), we have defined the cycle of the support
\( S_{T,n} \) on the torus \( T_{qp} \), eq.(\ref{e:support_n}). The last
result eq.(\ref{e:result}), is that the Chern index \( C_{n}=I(S_{n}) \) is
the homotopy number of this cycle (in the \( q \) direction). (We will now
drop the \( T \) suffix in \( S_{n} \)).

\begin{figure}[h]
{\par\centering \resizebox*{0.7\columnwidth}{!}{\includegraphics{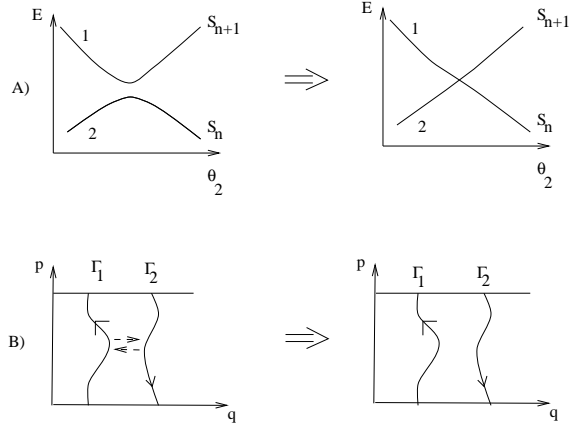}} \par}

\caption{\label{fig:croisement}The tunneling effect generates a splitting between energy
levels (A), and a jump of quasi-modes between trajectories (B). In order to
calculate the sum \protect\( \sum _{n=1}^{N}C_{n}\protect \), we have to remove
this splitting, as well as this jump.}
\end{figure}

We decide to define the sum of two or more consecutive cycles \( S_{n}+S_{n+1}+...+S_{n+a} \)
by removing the jumping due to tunneling at the crossings between two consecutive
cycles \( S_{i} \) and \( S_{i+1} \). This is illustrated on figure \ref{fig:croisement}.
The homotopy is then \( I(S_{n}+...+S_{n+a})=I(S_{n})+...+I(S_{n+a}) \), because
the removing is a local operation. 

\( \sum _{n=1}^{N}S_{n} \) is then the support of the semi-classical spectrum
\( \sigma _{sc} \) without tunneling, and from proposition (\ref{p:prop_cycles}),
it is composed of one cycle \( \Gamma _{nc} \) of homotopy 1 and many constant
cycles \( \Gamma _{c,i} \) of homotopy 0. This gives:
\[
\sum _{n=1}^{N}C_{n}=\sum _{n=1}^{N}I(S_{n})=I\left( \sum _{n=1}^{N}S_{n}\right) =+1.\]

We recover eq. (\ref{eq:sigmacn}). The total Chern index (\( +1 \)) is interpreted
here as the motion of the non contractible quasi-mode on the torus as \( \theta _{2} \)
is varying (in accordance with th.\ref{th:cylindre}, page \pageref{th:cylindre}).

\section{Numerical illustration}

We will illustrate the above results on the Hamiltonian eq.(\ref{e:exemple_H}).
The numerical calculations of the Chern indices have been done with the curvature
formula eq.(\ref{e:curvature}). For \( N=11 \) levels, the Chern indices are:

\vspace{0.3cm}
{\centering \begin{tabular}{|c|c|c|c|c|}
\hline 
\( C_{1\rightarrow 4} \) &
\( C_{5} \)&
\( C_{6} \)&
\( C_{7} \)&
\( C_{8\rightarrow 11} \)\\
\hline 
\hline 
\( 0 \)&
\( +1 \)&
\( -1 \)&
\( +1 \)&
\( 0 \)\\
\hline 
\end{tabular}\par}
\vspace{0.3cm}

Our analytical results have been obtained in the \( N\rightarrow \infty  \)
limit. Although, \( N=11 \) is not very high, the numerical results which follow
can well be interpreted in the semi-classical description of this paper.

The minimum value of \( N \) required to get correct semi-classical estimates,
can be evaluated in the following manner: the mesh provided by Planck cells
of magnitude \( 1/N \) has to be about so fine that all phase space structures
in a plot like figure \ref{fig:levels} can be resolved.

\begin{figure}[h]
{\par\centering \resizebox*{0.7\columnwidth}{!}{\includegraphics{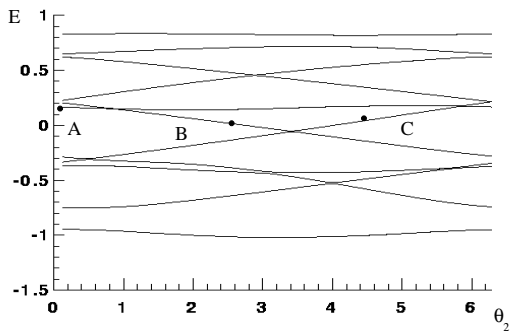}} \par}

\caption{\label{fig:bandes} Energy levels \protect\( E_{n}(\theta _{1},\theta _{2})\protect \)
for \protect\( n=1\rightarrow 11\protect \), of Hamiltonian eq.(\ref{e:exemple_H}),
as a function of \protect\( \theta _{2}\protect \). There is no degeneracy,
but the small splittings are not visible on the figure. Points A,B,C, correspond
to the Husimi representations of figure \ref{fig:husimi}.}
\end{figure}

Figure \ref{fig:bandes}, shows the energy levels \( E_{n}(\theta _{1},\theta _{2}) \)
for \( n=1\rightarrow 11 \), as a function of \( \theta _{2} \). The dependence
on \( \theta _{1} \) gives a width of the levels, but is very weak and not
visible on the figure. 

\begin{figure}[h]
{\par\centering \resizebox*{0.5\columnwidth}{!}{\includegraphics{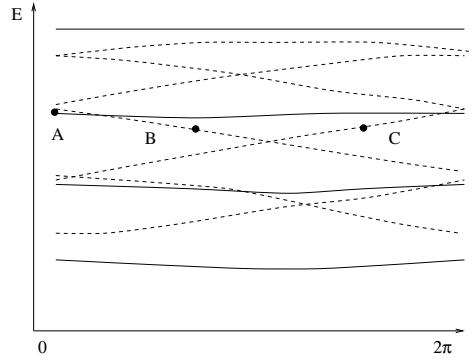}} \par}

\caption{\label{fig:bandes_sc} The semi-classical spectrum \protect\( \sigma _{sc}(\theta _{2})\protect \)
obtained from fig.(\ref{fig:bandes}). The four solid lines correspond to quasi-modes
\protect\( |\tilde{\psi }_{c}(\theta _{2})>\protect \) localized on contractible
trajectories. The dashed line corresponds to quasi-modes \protect\( |\tilde{\psi }_{nc}(\theta _{2})>\protect \)
localized on non-contractible trajectories. }
\end{figure}

In this energy spectrum, one can clearly distinguish two categories of energy
levels. These two categories are reproduced on figure \ref{fig:bandes_sc},
where we have artificially dropped the splittings. This figure corresponds to
the semi-classical spectrum \( \sigma _{sc}(\theta _{2}) \).

There are four energy levels (in solid line) which do not depend strongly on
\( \theta _{2} \) and have periodicity \( 2\pi  \). In the light of prop.
(\ref{p:prop_cycles}), these energy levels correspond to four quasi-modes \( |\tilde{\psi }_{c}(\theta _{2})> \)
localized on contractible trajectories. There is also one continuous energy
level \( E_{nc}(\theta _{2}) \) (in dashed line) which depends on \( \theta _{2} \)
and have periodicity \( 2\pi m \) with \( m=7 \). This energy level corresponds
to quasi-modes \( |\tilde{\psi }_{nc}(\theta _{2})> \) localized on non-contractible
trajectories. 

Figure \ref{fig:husimi}, shows the Husimi representation (see appendix (\ref{s:a1}))
of eigenstates \( |\varphi _{n}(\theta _{1},\theta _{2})> \) for the level
\( n=6 \), with \( \theta _{1}=0 \) , and three different values of \( \theta _{2} \).
These three eigen-states correspond to the points A,B,C on figures (\ref{fig:bandes})
and (\ref{fig:bandes_sc}). Looking at the classical trajectories on fig(\ref{fig:levels}),
one can clearly associate these quasi-modes respectively with the trajectories: 

(A): with a contractible trajectory (point D on fig \ref{fig:levels}), 

(B): a non-contractible trajectory of type \( (0,+1) \), 

(C): a non-contractible trajectory of type \( (0,-1) \). 

\begin{figure}[h]
{\par\centering \resizebox*{0.7\columnwidth}{!}{\includegraphics{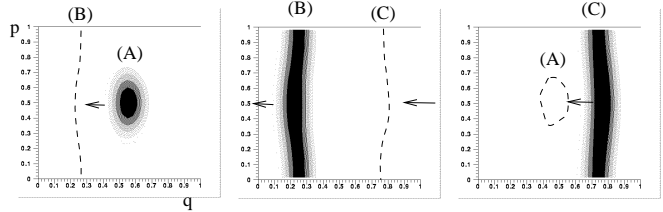}} \par}

\caption{\label{fig:husimi}The dark density are the Husimi representation of eigenstates
\protect\( |\varphi _{n}(\theta _{1},\theta _{2})>\protect \) for the band
\protect\( n=6\protect \) (see figure \ref{fig:bandes}), with \protect\( \theta _{1}=0\protect \)
, and three different values of \protect\( \theta _{2}\protect \). (A): \protect\( \theta _{2}=0\protect \);
\protect\( \quad \protect \)(B): \protect\( \theta _{2}=2,5\protect \),\protect\( \quad \protect \)(C):
\protect\( \theta _{2}=4,5\protect \). The dashed lines are the trajectories
where the quasi-modes will jump at the next crossing.}
\end{figure}

As \( \theta _{2} \) varies from \( 0 \) to \( 2\pi  \), the quantum state
\( |\varphi _{6}(\theta _{1},\theta _{2})> \) of level \( n=6 \) jumps. From
figure \ref{fig:bandes} and figure \ref{fig:bandes_sc} it is evident that
there are three (avoided) level crossings, where the quasi-mode jumps: 

Now from figure \ref{fig:husimi}, we can see that for the first crossing from
(A) to (B), the quasi-mode stays in the same cell (\( \Delta n_{1}=0 \)). For
the second crossing from (B) to (C) the quasi-mode changes by \( \Delta n_{1}=-1 \)
cell. For the third crossing from (C) to (A), the quasi-mode stays in the same
cell (\( \Delta n_{1}=0 \)). This gives a total change of \( \Delta n_{1}=-1 \)
cell and a homotopy \( I_{6}=-1 \) for this sequence of quasi-modes. Accordingly,
we have \( C_{6}=I_{6}=-1 \). Our result eq.(\ref{e:result}) is well verified
here.

\section{The Chern indices of the spectrum in a generic situation \label{s:generic}}

From the above results, we give here the precise values of Chern indices for
a simple and generic Hamiltonian (i.e. stable under perturbations) corresponding
to the numerical example eq.(\ref{e:exemple_H}) and explain how to read them
from the Reeb graph (fig. \ref{fig:bifurc}).

To simplify the discussion, we suppose in this section that we deal with a Hamiltonian
\( H \) such that the curve representing the non contractible trajectories
in the Reeb graph, has only one maximum and minimum, like the dashed curve on
fig. \ref{fig:bifurc}. We suppose moreover that in the energy range of this
curve, there is only one family of contractible trajectories, like the solid
line (C-D) on fig. \ref{fig:bifurc}.

\begin{figure}[h]
{\par\centering \resizebox*{0.8\columnwidth}{!}{\includegraphics{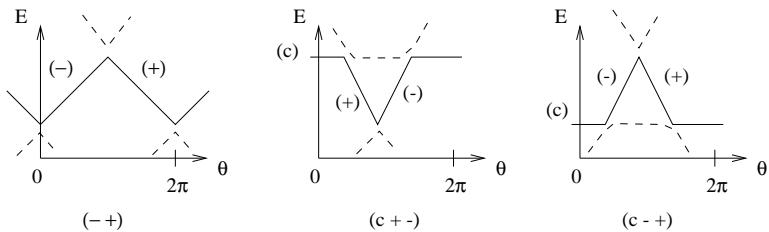}} \par}

\caption{\label{fig:type_bandes}The three possible band structures (solid lines), made
with a non contractible trajectory of type \protect\( (0,\pm 1)\protect \)
and a contractible trajectory \protect\( (c)\protect \). The dashed lines are
the upper and lower bands.}
\end{figure}

In the range of energy of the non-contractible trajectories, each band of the
spectrum is made with one of the three following sequences of quasi-modes, see
figure (\ref{fig:type_bandes}):

\begin{itemize}
\item The band is made with the two non contractible trajectories of type \( (0,\pm 1) \).
This gives the sequence: \( S=(+-) \).
\item The band has also a quasi-mode on the contractible trajectory \( (c) \). This
gives two possible sequences: \( S=(c+-) \) or \( S=(c-+) \). 
\end{itemize}

To simplify the discussion, we exclude sequences as \( (c+c-) \) or \( (c-c+) \).
(These sequences do not appear if the energy intervals between successive energy
levels of contractible quasi-modes are large enough).

It is clear from figure \ref{fig:type_bandes} that if the band \( n \) has
the sequence \( S_{n}=(c+-) \), then the band \( n+1 \) has necessary the
sequence \( S_{n+1}=(c-+) \), and conversely. These sequences come always by
pair.

To illustrate this from our numerical example, we can see on figure (\ref{fig:bandes})
and figure (\ref{fig:bandes_sc}), that the bands 3 to 9 have respectively the
sequences: \( S_{3}=(c+-) \), \( S_{4}=(c-+) \) ,\( S_{5}=(+-) \) ,\( S_{6}=(c+-) \)
,\( S_{7}=(c-+) \) ,\( S_{8}=(+-) \), \( S_{9}=(+-) \).

Now we saw in prop. (\ref{prop:chern_homotopy}) that a non zero Chern index
results from the jumping of neighboring quasi-modes on the cylinder, via non
contractible trajectories. We saw also that a given quasi-mode \( |\tilde{\psi }> \)
jumps by tunneling effect to the nearest quasi-mode of the same energy located
on a non contractible trajectory. It is therefore important to determine in
which direction (right or left) this jump occurs: we have to complete the Reeb
graph (fig \ref{fig:bifurc}), by giving the direction of the shortest jump
between any two trajectory. For that purpose, it is sufficient to give the location
of the separatrix between left-jumps and right-jumps. 

\begin{figure}[h]
{\par\centering \resizebox*{0.8\columnwidth}{!}{\includegraphics{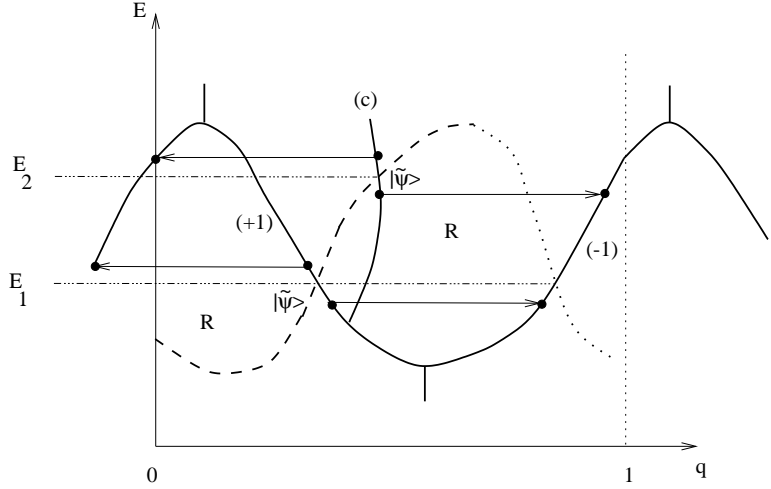}} \par}

\caption{\label{fig:tunnelling}Example for the jumping of a quasi-mode \protect\( |\tilde{\psi }>\protect \)
to the nearest non contractible trajectory (of type \protect\( (0,-1)\protect \)).
The solid curve is the Reeb graph as in fig. (\ref{fig:bifurc}). The dotted
and dashed lines represent the separatrix for these attraction basins. }
\end{figure}

On figure \ref{fig:tunnelling}, we have reproduced the Reeb graph (the solid
curve), and we have drawn (dashed line and dotted line) the possible location
of this separatrix. The effect of this separatrix (dashed line) is shown with
the plot of the jumps from one particular quasi-mode \( |\tilde{\psi }> \)
to a non-contractible trajectory of type \( (-1) \) at two different energies,
on both sides of the separatrix. The separatrix for the jump from/to a non contractible
of type \( (+1) \) is in dotted line. 

\( E_{1} \) is the energy where this separatrix crosses the non-contractible
trajectories family. That the energy \( E_{1} \) is the same for the dashed
and the dotted line comes simply from the reciprocity property of the tunneling
``geodesic''.

What matters in fact is only the intersection of this separatrix with the Reeb
graph. The picture \ref{fig:tunnelling} is qualitative here, but quantitative
calculations of the tunneling effect would precise the energy of these intersections. 

\begin{figure}[h]
{\par\centering \resizebox*{0.8\columnwidth}{!}{\includegraphics{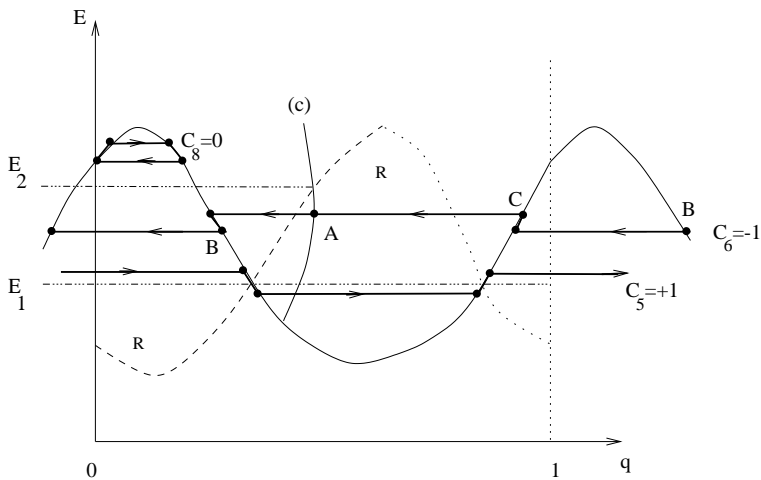}} \par}

\caption{\label{fig:general_sit}From knowledge of the sequences \protect\( S_{5},S_{6},S_{8}\protect \),
their homotopy and the Chern indices \protect\( C_{5}=+1\protect \), \protect\( C_{6}=-1\protect \)
and \protect\( C_{8}=0\protect \) are graphically computed. \protect\( C_{5}=+1\protect \)
is associated with a band of energy \protect\( E_{1}\protect \). Note that
\protect\( C_{6}=-1\protect \) is non zero thanks to the quasi-mode (A) on
the contractible trajectory (c) in region R.}
\end{figure}

Now, by combining figures \ref{fig:tunnelling} and \ref{fig:type_bandes} together
with rule eq.(\ref{e:result}), it is easy to compute the Chern index of the
three kind of bands. This is done on figure \ref{fig:general_sit} for the sequences
\( S_{5},S_{6},S_{8} \), and we infer the general result:

\begin{itemize}
\item If \( E_{1} \) is in the range of a band of type \( (+-) \) then its Chern
index is \( C_{(+-)}=+1 \), otherwise \( C_{(+-)}=0 \).
\item If the trajectory \( (c) \) is in the region \( R \) (of fig. \ref{fig:tunnelling})
then \( C_{(c+-)}=-1 \) (respect. \( +1 \)), if the band has energy greater
than \( E_{1} \) (respect. lower). In the other regions, \( C_{(c+-)}=0 \).
\item If the trajectory \( (c) \) is in the region \( R \) then \( C_{(c-+)}=+1 \)
(respect. \( -1 \)), if the band has energy greater than \( E_{1} \) (respect.
lower). In the other regions, \( C_{(c-+)}=0 \).
\end{itemize}
This allows us to interpret the Chern indices of our numerical example: the
band \( S_{5}=(+-) \) with \( C_{5}=+1 \) corresponds to the energy \( E_{1} \).
Above it, the two bands \( S_{6} \),\( S_{7} \) with Chern indices \( C_{6}=-1 \),
\( C_{7}=+1 \) form a pair, thanks to the presence of the contractible trajectory
\( (c) \) in region \( R \).

As a general result, we conclude that for any choice of Hamiltonian \( H \)
with a similar Reeb graph, there is a unique band in the middle of the spectrum,
including energy \( E_{1} \), with Chern index \( +1 \). Above this energy
(for \( E_{1}<E<E_{2} \)), the Chern indices come by pairs \( (-1,+1) \) ,
and under the energy \( E_{1} \), the Chern indices come by pairs \( (+1,-1) \).

Of course for another appearance of the Reeb graph, the rules could be different.
In particular, the Chern indices can reach higher values if the non-contractible
curve of the Reeb graph has more than one maximum. We do not know if there a
simple and explicit expression of the Chern indices in the general case.

\section{Chern index and Hall conductivity}

It is well known \cite{iqhe2} that the Chern index \( C \) of a band is related
to the transverse Integer Quantum Hall conductivity \( \sigma _{xy} \) by the
relation eq.(\ref{e:iqhe}). The usual demonstration of this formula starts
from the Kubo linear response formula for \( \sigma _{xy} \), and expresses
it as the curvature integral for \( C \), eq.(\ref{e:curvature}). In this
section we want to demonstrate it from an other point of view, by relating \( C \)
with the motion of a wave packet on the plane \( T_{qp} \). This result seems
more intuitive.

Consider a given energy band \( n \) of the spectrum eq.(\ref{e:spectre}).
For each \( (\theta _{1},\theta _{2}) \) there is one eigenvector \( |\varphi _{n}(\theta _{1},\theta _{2})> \)
defined up to a multiplicative constant. Suppose that we have a \( C^{\infty } \)
function \( (\theta _{1},\theta _{2})\rightarrow |\psi (\theta _{1},\theta _{2})>\in \mathcal{H}_{T}(\theta _{1},\theta _{2}) \),
where \( |\psi (\theta _{1},\theta _{2})> \) is proportional to \( |\varphi _{n}(\theta _{1},\theta _{2})> \),
eq.(\ref{e:spectre}). Note that if \( C_{n}\neq 0 \), this function must be
zero at some values of \( (\theta _{1},\theta _{2}) \).

Consider now 
\[
|\psi >=\int \int d\theta _{1}d\theta _{2}|\psi (\theta _{1},\theta _{2})>.\]

Then \( |\psi >\in \mathcal{H}_{P}=L^{2}(I\! \! R) \). We can say that \( |\psi > \)
is a state ``localized'' on the plane \( T_{qp} \) constructed from the band
\( n \), like Wannier states in Solid State Physics. (Conversely, given \( |\Phi >\in L^{2}(I\! \! R) \),
we can project it on the spectral band \( n \) and obtain such \( |\psi > \).) 

The state \( |\psi > \) can be translated by \( (n_{1},n_{2}) \) cells on
the plane by:
\[
|\psi _{n_{1},n_{2}}>=T_{Q}^{n_{1}}T_{P}^{n_{2}}|\psi >=\int \int d\theta _{1}d\theta _{2}e^{in_{1}\theta _{1}+in_{2}\theta _{2}}|\psi (\theta _{1},\theta _{2})>.\]

Conversely:
\[
|\psi (\theta _{1},\theta _{2})>=\sum _{n_{1},n_{2}\in Z\! \! \! Z^{2}}e^{-in_{1}\theta _{1}-in_{2}\theta _{2}}|\psi _{n_{1},n_{2}}>.\]

These simple relations can be generalized by the following result:

Consider an other state constructed from the band \( n \):
\begin{equation}
\label{e:phase}
|\phi >=\int \int d\theta _{1}d\theta _{2}e^{if(\theta _{1},\theta _{2})}|\psi (\theta _{1},\theta _{2})>,
\end{equation}

where \( f \) is a continuous and periodic function such that:
\begin{eqnarray*}
f(\theta _{1}+2\pi ,\theta _{2})=f(\theta _{1},\theta _{2})+N_{1}2\pi  &  & \\
f(\theta _{1},\theta _{2}+2\pi )=f(\theta _{1},\theta _{2})+N_{2}2\pi , &  & 
\end{eqnarray*}
(\( N_{1},N_{2}\in Z\! \! \! Z^{2} \) are topological integers which characterized
the homotopy of the function \( f \)). 

Let us define:
\begin{eqnarray*}
<n_{1}>=\sum _{n_{1},n_{2}}n_{1}.\left| <\psi _{n_{1},n_{2}}|\phi >\right| ^{2} &  & \\
<n_{2}>=\sum _{n_{1},n_{2}}n_{2}.\left| <\psi _{n_{1},n_{2}}|\phi >\right| ^{2}, &  & 
\end{eqnarray*}
which give the mean position of the state \( |\phi > \)on the plane \( T_{qp} \)
relatively to the localized states \( |\psi _{n_{1},n_{2}}> \). It is easy
to prove that:
\begin{eqnarray*}
<n_{1}>=N_{1} &  & \\
<n_{2}>=N_{2}, &  & 
\end{eqnarray*}
the mean position of \( |\phi > \)is then ``quantized''.

(This is a simple property of Fourier series: if \( g(\theta )=\sum _{n}c_{n}e^{in\theta }=e^{if(\theta )} \)
is periodic and \( f(2\pi )=f(0)+N2\pi  \) then \( N=\sum _{n}n|c_{n}|^{2} \).
Another way to say it is: \( \sum _{n}n|c_{n}|^{2}=<g|\hat{p}_{\theta }|g>=\frac{1}{2\pi }\int _{0}^{2\pi }f'd\theta =N \),
with the current operator \( \hat{p}_{\theta }=\frac{1}{i}d/d\theta  \).).

We now give some consequences of this result on the ``quantized mean position''.

Now let us first look at the time evolution of \( |\psi _{0,0}> \):
\[
|\psi (t)>=e^{-iHt/\hbar }|\psi _{0,0}>=\int \int d\theta _{1}d\theta _{2}e^{-iE(\theta _{1},\theta _{2})t/\hbar }|\psi (\theta _{1},\theta _{2})>.\]
Comparing to eq.(\ref{e:phase}), the phase function is the dynamical phase
\( f(\theta _{1},\theta _{2})=-E(\theta _{1},\theta _{2})t/\hbar  \) of homotopy
type \( N_{1}=N_{2}=0 \) . (because by continuous homotopy \( t\rightarrow 0 \),
\( f \) is mapped to \( 0 \)). This means that during its evolution, the quantum
state \( |\psi (t)> \)spreads over the plane, but its mean position \( <n_{1}> \),\( <n_{2}> \)
stays zero.

Let us suppose now that \( \theta _{1},\theta _{2} \) are no more ``good quantum
numbers'', and that there is a slow drift. In the model of bi-dimensional electrons
of the introduction, this drift is the adiabatic motion due to the low electrical
field \( E_{x} \). Consider for example:

\begin{eqnarray*}
\theta _{1}(t)=-\omega _{1}t &  & \\
\theta _{2}(t)=\theta _{2}(0). &  & 
\end{eqnarray*}

After one period \( T=2\pi /\omega _{1} \), the evolution gives
\[
e^{-iHT/\hbar }|\psi (\theta _{1},\theta _{2})>=\exp \left( i\phi _{D}(\theta _{1},\theta _{2})+i\phi _{B}(\theta _{2})\right) |\psi (\theta _{1},\theta _{2})>,\]
where \( \phi _{D}(\theta _{1},\theta _{2}) \) is the dynamical phase of homotopy
type \( N_{1}=N_{2}=0 \) and \( \phi _{B}(\theta _{1},\theta _{2}) \) is the
Berry's phase of the path \( \theta _{1}(t) \) and doesn't depend on \( \theta _{1} \)
henceforth. By homotopy deformation, \( \phi _{B}(\theta _{2})\equiv 2\pi C\theta _{2} \)
and is therefore of type \( N_{1}=0 \), \( N_{2}=C \). This means that after
one period \( T \), the mean position of the quantum state \( |\psi (t)> \)has
been changed by \( \delta <n_{1}>=0 \), \( \delta <n_{2}>=C \) (in cells units).

This ``quantized'' velocity \( V_{2}=\delta <n_{2}>/T \) is responsible for
the Integer Hall conductivity. Indeed, in the Harper Model \cite{iqhe2} of
non interacting bi-dimensional \( (x,y) \) electrons in a bi-periodic potential
of period \( (X,Y) \), and a perpendicular high magnetic field \( B_{z} \),
with a low electrical field along \( E_{x} \), the adiabatic hypothesis gives:
\[
\theta _{1}(t)=-\frac{eX}{\hbar }E_{x}t.\]
We have obtained that the quantized velocity of one electron quantum state is
\( V_{y}=(CY)/T \). For a filled band, the electronic density is one electron
per cell: \( \rho =1/(XY) \). The electronic density current of a filled band
is then \( j_{y}=\rho eV_{y}=\frac{e^{2}}{h}C.E_{x} \) which leads to eq.(\ref{e:iqhe}).

\section{Conclusion}

By using quasi-modes, we have obtained a semi-classical description of band
eigen-states of a generic quantum Hamiltonian on the torus phase-space. By using
general results exposed in the appendices, relating the topological Chern index
of a band, with the localization properties of the quantum states, we have been
able to express the Chern index of each band in terms of the motion of the quasi-modes
on the classical phase space. 

In section 7, we have shown how the Chern indices of the whole spectrum can
be graphically computed from the classical Reeb graph. There is an interesting
question which can be raised from our work: conversely, from a given sequence
\( (C_{n})_{n} \) of integers such that \( \sum _{n}C_{n}=+1 \), the question
would be to determine the potential \( V(x,y) \) whose band spectrum possess
this precise sequence of Chern indices.

Currently one tries to observe experimental signatures of the Harper spectrum
(Landau level substructures) in superlattices with periods of about \( 100 \)
nm on GA-As-AlGaAs heterojunctions (\cite{gudmundsson},\cite{schlosser}).
In these experiments, the Chern index gives the quantum Hall conductivity \( \sigma _{xy} \).
In this paper we have shown simple correspondances between the Chern index and
the classical trajectories of the electrons. This correspondances could be observed
in experiments by measuring the variations of the integer Hall conductivity,
in relation with the potential \( V(x,y) \) of the supperlattice, which may
be created by an external electrostatic grid.

\section{Acknowledgments}

Discussions with B. Parisse and Y. Colin de Verdiere are gratefully acknowledged.

\appendix

\section{Bargmann and Husimi representation\label{s:a1}}

In this appendix, we recall well known results on Bargmann and Husimi representation
on the torus phase space (\cite{voros1}).

We have seen previously that the space \( \mathcal{H}_{T} \) is not a subspace
of \( L^{2}(I\! \! R) \). For functions belonging to \( \mathcal{H}_{T} \),
it will be more useful to introduce a phase space representation of the quantum
states, called the Bargmann representation (\cite{bargmann}).

Consider a quantum state \( |\psi >\in L^{2}(I\! \! R) \). In order to characterize
the localization of \( |\psi > \) in the phase space near the point \( (q,p) \),
we first construct a Gaussian wave packet \( |qp> \) (coherent state) defined
in the \( x \)-representation by: 
\[
<x|qp>=\left( \frac{1}{\pi \hbar }\right) ^{1/4}\exp (\frac{i}{\hbar }px)\exp (-\frac{(x-q)^{2}}{2\hbar }).\]
 The notation \( |qp> \) recalls that the coherent state is localized (in the
semi-classical limit) at the point \( (q,p) \) of the phase space.

The Husimi distribution of a state \( |\psi > \) is defined over the phase
space by: 
\[
h_{\psi }(q,p)=|<qp|\psi >|^{2},\]
 and for \( \varphi \in L^{2}(I\! \! R) \), we have: 
\begin{equation}
\label{eq:coherentL2}
\int |\varphi (x)|^{2}\, \, dx=\int \int |<qp|\varphi >|^{2}\, \, \frac{dq\, \, dp}{2\pi \hbar }.
\end{equation}

To characterize the functions of \( L^{2}(T^{*}I\! \! R) \) which are \( (q,p) \)
representations of a state, it is more convenient to introduce a complex representation
of the phase space \( z=\frac{1}{\sqrt{2\hbar }}(q+ip) \). Another (proportional)
expression of the coherent state is then: 
\[
|z>=\exp (za^{+})|0>,\]
 with \( |0> \) being the fundamental of the harmonic oscillator \( H_{0}=\hat{q}^{2}+\hat{p}^{2}, \)
and \( a^{+} \) being the associated creation operator. Indeed: 
\[
|qp>=\exp (i\frac{qp}{2\hbar }-\frac{q^{2}+p^{2}}{4\hbar })|z>.\]
 The following anti-holomorphic function of \( z \) is called the Bargmann
distribution of \( \psi  \): 
\[
b_{\psi }(z)=<z|\psi >.\]

Clearly, we have 
\[
h_{\psi }(q,p)=\left| b_{\psi }(z)\right| ^{2}e^{-\frac{q^{2}+p^{2}}{2h}},\]
 hence the zeroes of the function \( h_{\psi }(q,p) \) are those of the holomorphic
function \( b_{\psi }(z) \), which are localized zeroes in the phase space.
Moreover, (\ref{eq:coherentL2}) implies that \( \psi \in L^{2}(I\! \! R) \)
if and only if \( b_{\psi }\in L^{2}(I\! \! \! \! C,e^{-|z|^{2}/h}) \) and
\( b_{\psi } \) is anti-holomorphic.

The same definitions can be applied for a state \( |\psi >\in \mathcal{H}_{T}(\theta _{1},\theta _{2}) \).
The corresponding Bargmann function is a theta-function \cite{zero2} and the
Husimi distribution is bi-periodic in \( (q,p) \), hence is well defined on
the torus \( T_{qp} \). 

We need the following properties concerning the Bargmann representation \( b_{\psi }(z)=<z|\psi > \)
of a quantum state \( |\psi > \) on the torus (\( |z> \) is a coherent state
on the torus \cite{perelomov1}:

If \( |\psi >\in \mathcal{H}_{T}(\theta _{1},\theta _{2}) \), the Bargmann
function \( b_{\psi }(z) \) is (a theta-function) anti-holomorphic with respect
to \( z \), and with \( N \) zeros \( Z=(z_{1}\dot{,}\ldots ,z_{N}) \) in
the cell \( [0,1]\times [0,1] \). These zeros are constrained by (\cite{zero2},
this is ``Abel theorem'', and ``Jacobi inversion theorem'', see \cite{harris1}
p.235):

\begin{eqnarray}
\sum _{i=1}^{N}q_{i}=\frac{\theta _{2}}{2\pi } &  & \label{e:constraint} \\
\sum _{i=1}^{N}p_{i}=\frac{\theta _{1}}{2\pi }. &  & \nonumber 
\end{eqnarray}

with \( z_{i}=q_{i}+ip_{i} \). Conversely, such a collection of \( N \) zeros
define a state \( |\psi >\in \mathcal{H}_{T}(\theta _{1},\theta _{2}) \), unique
up to a multiplicative constant.

Note that since the \( N \) zeroes are constrained to have a fixed sum, we
get the right dimension \( N-1+1 \) for the Hilbert space \( \mathcal{H}_{T}(\theta _{1},\theta _{2}) \).

\section{Calculation of the Chern index in special cases\label{s:a2}}

For simplicity, we will note \( \theta =(\theta _{1},\theta _{2})\in I\! \! R^{2} \).

We will compute the Chern index of a line bundle in particular cases. We will
always consider a line sub-bundle of \( \mathcal{H}_{T}(\theta )\rightarrow T_{\theta } \).
Such a line bundle \( F \) is characterized by giving local section \( \theta \rightarrow |\psi _{i}(\theta )>\in \mathcal{H}_{T}(\theta ) \)
over open sets \( U_{i} \) which cover \( T_{\theta } \).

From appendix \ref{s:a1}, the line bundle \( F \) is also characterized by
the function \( \theta \rightarrow Z(\theta ) \) , where \( Z(\theta ) \)
is the set of \( N \) non-ordered zeroes of the Bargmann functions \( b_{\psi (\theta )}(z) \).

\subsection{Theorem on homotopy invariance}

\begin{thm}
\emph{\label{th:homotopy}}. Consider two complex line bundles over \( T_{\theta } \)
with Chern index \( C \) and \( C' \) defined by local sections \( |\psi _{i}(\theta )> \)and
\( |\psi _{i}'(\theta )> \)on open sets \( U_{i} \). Suppose moreover that
\[
<\psi _{i}(\theta )|\psi _{i}'(\theta )>\neq 0,\quad \forall \theta ,\; \forall i,\]
 then \( C=C' \) .
\end{thm}
\begin{proof}
We can suppose that the sections are normalized \( (<\psi _{i}(\theta )|\psi _{i}(\theta )>=1) \).
On a set \( U_{i} \), and for a given \( \theta  \), define: \textsf{
\[
\rho _{i}(\theta )=\left| <\psi _{i}(\theta )|\psi _{i}'(\theta )>\right| \leq 1,\]
} and for fixed \textsf{\( \lambda \in [0,1] \):
\[
|\varphi _{i,\lambda }(\theta )>=|\psi _{i}(\theta )>.<\psi _{i}(\theta )|\psi _{i}'(\theta )>+\lambda \left( |\psi _{i}'(\theta )>-|\psi _{i}(\theta )><\psi _{i}(\theta )|\psi _{i}'(\theta )>\right) .\]
} In the theorem, we suppose that \( \rho _{i}(\theta )>0. \) Thus 
\[
<\psi _{i}'(\theta )|\varphi _{i,\lambda }(\theta )>=\rho ^{2}+\lambda (1-\rho ^{2})>0,\]
 and then \( |\varphi _{i,\lambda }(\theta )>\neq 0 \).

We now check that the sections \( |\varphi _{i,\lambda }(\theta )> \) define
a complex line bundle \( F_{\lambda } \). On an other set \( U_{j} \), if
\( |\psi _{j}(\theta )>=\exp (i\alpha )\; \dot{|}\psi _{i}(\theta )> \)and
\( |\psi _{j}'(\theta )>=\exp (i\beta )\; \dot{|}\psi _{i}'(\theta )> \) then
\( |\varphi _{j,\lambda }(\theta )>=\exp (i\beta )\; |\varphi _{i,\lambda }(\theta )> \)
is in the same line than \( |\varphi _{i,\lambda }(\theta )> \).

We have thus obtain a homotopic deformation \( F_{\lambda } \), \( \lambda \in [0,1] \)
between the line bundle \( F_{0} \) and \( F_{1} \). They have thus the same
Chern index \( C=C_{\lambda }=C' \) .
\end{proof}
\begin{thm}
\label{eq:chern_general}More generally (\cite{harris1},p. 141) if we suppose
that 
\[
\theta \rightarrow <\psi _{i}(\theta )|\psi _{i}'(\theta )>,\]
 has zeros in \( \theta ^{*} \) with index \( \iota (\theta ^{*})=\pm 1 \)
,then 

\begin{equation}
\label{e:chern}
C=C'+\sum _{\textrm{zeros }\theta ^{*}}\iota (\theta ^{*}).
\end{equation}
 
\end{thm}

\subsection{The Chern index from the zeroes of the Bargmann function}

Let \( F \) be a line bundle as above.

\begin{itemize}
\item Suppose that there exists some point \( z_{0}\in T_{qp} \) of phase space,
such that 
\[
\forall \theta \in T_{\theta },\quad b_{\psi (\theta )}(z_{0})\neq 0,\]
 (equivalently, this means that \( z_{0}\notin Z(\theta ) \), for all \( \theta  \)).
Then we can select a vector \( |\psi (\theta )>\in H_{N}(\theta ) \) in each
fiber such that \( \arg \left( b_{\psi (\theta )}(z_{0})\right) =0 \). This
gives a non-vanishing global section of the bundle. This is also a global frame,
hence the bundle is trivial, and \( C=0 \). 
\item More generally, define 
\[
N(z_{0})=\left\{ \theta \in T_{\theta }\quad /\quad z_{0}\in Z(\theta )\right\} ,\]
 then 
\begin{equation}
\label{e:chern2}
C=\sum _{\theta \in N(z_{0})}(\pm 1),
\end{equation}
 where the sign \( \pm 1 \) corresponds to the local orientation of the mapping
\( z_{i} \) at \( \theta  \), where \( Z=\left\{ z_{1},\ldots ,z_{N}\right\}  \)
and \( z_{i}(\theta )=z_{0} \). This can be deduced directly from eq.(\ref{e:chern}),
with \( |\psi '>=|z_{0}> \). 
\end{itemize}

\subsection{Example of a line bundle with a given Chern index \protect\( C\protect \)}

We now construct explicit examples of such a bundle.

\begin{itemize}
\item Suppose that \( N\geq 2 \). We define a line bundle \( F_{0} \), by specifying
the zeros of the Bargmann representation of a section, with the notations of
eq.(\ref{e:constraint}):
\begin{eqnarray}
q_{1}=\frac{\theta _{2}}{2\pi },\quad p_{1}=0 &  & \label{e:f0} \\
q_{2}=0,\quad p_{2}=\frac{\theta _{1}}{2\pi } &  & \nonumber \\
q_{i}=0,\quad p_{i}=0\textrm{ for }3\leq i\leq N. &  & \nonumber 
\end{eqnarray}
 The constraint eq.(\ref{e:constraint}) is easily verified. If we choose \( z_{0}=(q_{0}+ip_{0}) \)
with \( q_{0}=p_{0}=0.5 \), \( z_{0}\notin Z(\theta ) \) and from eq. (\ref{e:chern2}),
we deduce that the bundle \( F_{0} \) is trivial, with Chern index \( C=0 \). 
\item More generally, consider the bundle \( F_{C} \) defined by 
\begin{eqnarray*}
q_{1}=(1-C)\frac{\theta _{1}}{2\pi },\quad p_{1}=0 &  & \\
q_{2}=C\frac{\theta _{2}}{2\pi },\quad p_{2}=\frac{\theta _{1}}{2\pi } &  & \\
q_{i}=0,\quad p_{i}=0 & \textrm{for }3\leq i\leq N. & 
\end{eqnarray*}
 The constraint eq.\ref{e:constraint} is verified. We can apply eq. (\ref{e:chern2}),
to calculate the Chern index \( C \) . The set \( N(z_{0}) \) is given by
\[
\theta _{1}=\pi \; [2\pi ],\quad \theta _{2}=\frac{\pi }{C}\; [2\pi /C],\]
 and we obtain that the Chern index is \( C \). 
\end{itemize}

\subsection{Chern index for a moving coherent state }

\begin{thm}
\textbf{\label{th:ec}}Suppose \( N\geq 2 \). Consider the line bundle \( F_{z} \)
defined by the local section \( |\psi (\theta )>=|z_{\theta }>\in \mathcal{H}_{T}(\theta ) \)
with \( z_{\theta }=q_{\theta }+ip_{\theta } \), and 
\begin{equation}
\label{e:pq}
\left( \begin{array}{c}
q_{\theta }\\
p_{\theta }
\end{array}\right) =\left( \begin{array}{cc}
n_{11} & n_{12}\\
n_{21} & n_{22}
\end{array}\right) \left( \begin{array}{c}
\theta _{1}/2\pi \\
\theta _{2}/2\pi 
\end{array}\right) ,
\end{equation}

(This means that the coherent state \( |z_{\theta }> \) is moving over the
cells as \( \theta  \) is varying.)

The Chern index of this bundle is then
\begin{equation}
\label{e:chern-ec}
C=N.\left| \begin{array}{cc}
n_{11} & n_{12}\\
n_{21} & n_{22}
\end{array}\right| +n_{21}+n_{12}.
\end{equation}
 
\end{thm}
\begin{proof}
Consider the line bundle \( F_{0} \) defined by the zeroes eq.(\ref{e:f0}),
with local section noted \( |\varphi _{i}(\theta )> \) on each open set \( U_{i}\subset T_{\theta } \).
This bundle has \( C=0 \) Chern index. Consider the function 
\[
f:\theta \rightarrow <\varphi _{i}(\theta )|z_{\theta }>.\]
 From eq.(\ref{e:chern}), the Chern index \( C \) of the bundle \( F_{z} \)
is 
\[
C=\sum _{\theta ^{*}\textrm{zeroes of f}}\iota (\theta ^{*}).\]

The zeroes of \( f \) are given by: 
\begin{eqnarray*}
<\varphi _{i}(\theta )|z_{\theta }>=0 & \Leftrightarrow \; z_{\theta }\textrm{ is a zero of the section }|\varphi _{i}(\theta )> & \\
\textrm{ } & \Leftrightarrow \left\{ \begin{array}{c}
q_{\theta }=q_{i}\\
p_{\theta }=p_{i}
\end{array}\right. ,i=1,\ldots ,N. & \\
 &  & 
\end{eqnarray*}
 this gives 
\begin{eqnarray*}
\left\{ \begin{array}{c}
n_{11}\frac{\theta _{1}}{2\pi }+n_{12}\frac{\theta _{2}}{2\pi }\\
n_{21}\frac{\theta _{1}}{2\pi }+n_{22}\frac{\theta _{2}}{2\pi }
\end{array}\right. \textrm{ }=\left\{ \begin{array}{c}
\frac{\theta _{2}}{2\pi }\\
0
\end{array}\right. \textrm{or }=\left\{ \begin{array}{c}
0\\
\frac{\theta _{1}}{2\pi }
\end{array}\right.  & \textrm{or }=\left\{ \begin{array}{c}
0\\
0
\end{array}\textrm{for }i=3,\ldots ,N.\right.  & \\
 &  & 
\end{eqnarray*}

We deduce from eq.(\ref{eq:chern_general}) that
\begin{eqnarray*}
C= & \left| \begin{array}{cc}
n_{11} & n_{12}-1\\
n_{21} & n_{22}
\end{array}\right| +\left| \begin{array}{cc}
n_{11} & n_{12}\\
n_{21}-1 & n_{22}
\end{array}\right| +(N-2)\left| \begin{array}{cc}
n_{11} & n_{12}\\
n_{21} & n_{22}
\end{array}\right|  & \\
= & N\left| \begin{array}{cc}
n_{11} & n_{12}\\
n_{21} & n_{22}
\end{array}\right| +n_{21}+n_{12}. & 
\end{eqnarray*}
 
\end{proof}

\subsection{Bundle constructed from periodic motion of states on the plane}

A natural question is : is it possible to generalize the theorem \ref{th:ec},
for the periodic motion of arbitrary states in \( L^{2}(I\! \! R) \) rather
than only coherent states?

There is a first result:

\begin{thm}
\textbf{\label{th:bundle}}Let \( S:\theta \rightarrow |\psi (\theta )>\in L^{2}(I\! \! R) \)
for \( \theta \in I\! \! R^{2} \) be an arbitrary mapping such that :
\begin{eqnarray}
\forall \theta ,\quad |\psi (\theta )>\neq 0 &  & \label{e:plan_periodic} \\
|\psi (\theta _{1}+2\pi ,\theta _{2})>=T_{Q}^{n_{11}}T_{P}^{n_{12}}|\psi (\theta )> &  & \nonumber \\
|\psi (\theta _{1},\theta _{2}+2\pi )>=T_{Q}^{n_{21}}T_{P}^{n_{22}}|\psi (\theta )> & . & \nonumber 
\end{eqnarray}
 with \( (n_{11},n_{12},n_{21},n_{22})\in Z\! \! \! Z^{4} \) which characterizes
the periodic motion of the states \( |\psi (\theta )> \) as \( \theta  \)
is varied. 

For a contractible open subset \( U \) of \( T_{\theta } \), define 
\[
|\tilde{\psi }(\theta )>=P_{\theta _{1}}P_{\theta _{2}}(|\psi (\theta )>)\in \mathcal{H}_{T}(\theta ).\]
 Then if \( N\geq 3 \), and for a generic map \( S \), \( |\tilde{\psi }(\theta )> \)
gives a local section of a well defined line bundle over \( T_{\theta } \),
which can be noted also \( S \).
\end{thm}
\begin{proof}
For every \( \theta  \), the line bundle is defined by the vector \( |\tilde{\psi }(\theta )> \).
We have therefore to check that \( |\tilde{\psi }(\theta )>\neq 0 \) , and
that this line is periodic with respect to \( \theta  \).

Because the space \( \mathcal{H}_{T}(\theta ) \) is \( N \) dimensional, the
condition \( |\tilde{\psi }(\theta )>=0 \) is \( N \)-dimensional. As soon
as \( N>2 \), and for \( \theta \in [0,2\pi ]^{2} \)the condition \( |\tilde{\psi }(\theta )>=0 \)
cannot be satisfied generically. Now 
\begin{eqnarray*}
|\tilde{\psi }(\theta _{1}+2\pi ,\theta _{2})>=P_{\theta _{1}+2\pi }(|\psi (\theta _{1}+2\pi )>) &  & \\
=P_{\theta _{1}}T_{Q}^{n_{11}}T_{P}^{n_{12}}|\psi (\theta _{1})> &  & \\
=\exp (in_{11}\theta _{1})\exp (in_{12}\theta _{2})|\tilde{\psi }(\theta )>. &  & 
\end{eqnarray*}
 So \( |\tilde{\psi }(\theta )> \) and \( |\tilde{\psi }(\theta _{1}+a2\pi ,\theta _{2}+b2\pi )> \)are
proportional and define the same line \( \forall (a,b)\in Z\! \! \! Z^{2} \).
\end{proof}
\begin{thm}
\textbf{\label{th:L2}}Suppose that \( N\geq 4 \). Let \( S:\theta \rightarrow |\psi (\theta )>\in L^{2}(I\! \! R) \)
be a generic mapping as defined in theorem \ref{th:bundle}a. Then the Chern
index of the line bundle \( S \) is given by formula (\ref{e:chern-ec}).
\end{thm}
\begin{proof}
We note \( S_{0}=S \), and \( S_{1}=F_{z} \) the line bundle defined in theorem
\ref{th:ec}. The proof consists in constructing a homotopic deformation \( S_{\lambda } \)
from \( S_{0} \) to \( S_{1} \) for \( \lambda \in [0,1] \). Because the
Chern index \( C_{\lambda } \) of \( S_{\lambda } \) is constant, we therefore
conclude that the Chern index of \( S \) is given by formula (\ref{e:chern-ec}).

For \( \lambda \in [0,1] \), and \( \theta \in I\! \! R^{2} \), define 
\[
|\psi _{\lambda }(\theta )>=\lambda |\psi (\theta )>+(1-\lambda )|z(\theta )>\in L^{2}(I\! \! R),\]
 where \( |z(\theta )> \) is a standard coherent state on the plane, \( z(\theta )=q_{\theta }+ip_{\theta } \),
with \( q_{\theta },p_{\theta } \) given by eq. (\ref{e:pq}). We suppose moreover
that the phases are chosen such that eq. (\ref{e:plan_periodic}) holds for
\( |z(\theta )> \) and \( |\psi (\theta )> \).

Then \( |\psi _{\lambda }(\theta )>=0 \) if and only if \( \lambda =1/2 \)
and \( |\psi (\theta )>=-|z(\theta )> \). This last situation is non generic,
and if it occurs, we can just choose \( z(\theta )=(q_{\theta }+q_{0})+i(p_{\theta }+p_{0}) \)
with arbitrary \( q_{0},p_{0} \) to avoid this. So \( S_{\lambda }:\theta \rightarrow |\psi _{\lambda }(\theta )>\in L^{2}(I\! \! R) \)
fulfills theorem \ref{th:bundle}, and defines a line bundle for every \( \lambda  \).
Because there are three parameters \( \theta _{1},\theta _{2},\lambda  \) we
have to suppose now that \( N\geq 4 \) so that \( |\tilde{\psi }(\theta )>\neq 0 \).
\end{proof}

\subsection{Bundle constructed from a periodic motion on the cylinder}

In section \ref{s:quasi-modes}, we obtain the periodic motion of quasi-modes
on the cylinder \( C_{qp} \) phase-space. A slightly different result than
theorem \ref{th:L2} is then needed:

\begin{thm}
\textbf{\label{th:cylindre}}Suppose \( N\geq 4 \), and let 
\[
\theta _{2}\in I\! \! R\rightarrow |\psi _{C}(\theta _{2})>\in \mathcal{H}_{C}(\theta _{2}),\]
 be a generic continuous mapping such that 
\begin{eqnarray*}
\forall \theta _{2}\in I\! \! R,\qquad |\psi _{C}(\theta _{2})>\neq 0, &  & \\
|\psi _{C}(\theta _{2}+2\pi )>=T_{Q}^{I}|\psi _{C}(\theta _{2})>, &  & 
\end{eqnarray*}
 with \( I\in Z\! \! \! Z \).

For \( \theta =(\theta _{1},\theta _{2}) \) define 
\[
|\tilde{\psi }(\theta )>=P_{\theta _{1}}|\psi _{C}(\theta _{2})>\in \mathcal{H}_{T}(\theta ).\]
 Then \( \theta \rightarrow |\tilde{\psi }(\theta )>\in \mathcal{H}_{T}(\theta ) \)
is a local section of a well defined line bundle over \( T_{\theta } \), with
Chern index \( I \). 
\end{thm}
\begin{proof}
The proof is similar to that of theorem \ref{th:L2}.

The bundle is well defined because, first \( |\tilde{\psi }(\theta )>=0 \)
needs \( N \) conditions, generically not realized for arbitrary \( \theta _{1},\theta _{2}\in [0,2\pi ] \),
as soon as \( N>2 \).

Secondly, \( |\tilde{\psi }(\theta )> \) and \( |\tilde{\psi }(\theta _{1}+a2\pi ,\theta _{2}+b2\pi )>=P_{\theta _{1}}T_{Q}^{Ib}|\psi _{C}(\theta _{2})>=\exp (iIb\theta _{1})|\tilde{\psi }(\theta )> \)
are proportional and define therefore the same line.

In order to calculate the Chern index, consider the mapping 
\[
\theta _{2}\in I\! \! R\rightarrow |z(\theta _{2})>\in L^{2}(I\! \! R),\]
 where \( |z(\theta _{2})> \) is a standard coherent state on the plane, \( z(\theta _{2})=q_{\theta }+ip_{\theta } \),
and 
\begin{eqnarray*}
q_{\theta }=I.\theta _{2} &  & \\
p_{\theta }=p_{0}=constant, &  & 
\end{eqnarray*}
 we can therefore construct \( |z(\theta _{2})_{C}>=P_{\theta _{2}}|z(\theta _{2})>\in \mathcal{H}_{C}(\theta _{2}) \)
and \( |z(\theta _{1},\theta _{2})>=P_{\theta _{1}}|z(\theta _{2})_{C}>\in \mathcal{H}_{T}(\theta ) \).

From theorem \ref{th:ec}, \( \theta \rightarrow |z(\theta _{1},\theta _{2})> \)
define a line bundle with Chern index \( n_{12}=I \).

Consider now for \( \lambda \in [0,1] \) 
\[
\theta _{2}\rightarrow |\psi _{\lambda }(\theta _{2})>=\lambda |\psi (\theta _{2})>+(1-\lambda )|z(\theta _{2})_{C}>\in \mathcal{H}_{C}(\theta _{2}),\]
 and 
\[
S_{\lambda }:\theta =(\theta _{1},\theta _{2})\rightarrow |\tilde{\psi }_{\lambda }(\theta )>=P_{\theta _{1}}|\psi _{\lambda }(\theta _{2})>\in H_{T}(\theta ).\]
 Then for fixed \( \lambda  \), \( S_{\lambda } \) define a line bundle over
\( T_{\theta } \), because \( |\tilde{\psi }_{\lambda }(\theta )>=0 \) nor
\( |\psi _{\lambda }(\theta )>=0 \) are generic (cf discussion in proof of
theorem \ref{th:L2}), and there is also periodicity with respect to \( \theta  \).

The mapping \( \lambda \rightarrow S_{\lambda } \) is a homotopic deformation
from the line bundle \( \theta \rightarrow |\tilde{\psi }(\theta )> \) to the
line bundle \( \theta \rightarrow |z(\theta _{1},\theta _{2})> \). Their Chern
index is constant and equal to \( I \).

\newpage
\end{proof}
\bibliographystyle{unsrt}

\end{document}